\documentclass[preprint,12pt]{elsarticle}

\usepackage{graphicx}
\usepackage{amstext}
\usepackage{amsmath}

\begin{document}

\title{BerkeleyGW: A Massively Parallel Computer Package for the Calculation of the Quasiparticle and Optical Properties of Materials and Nanostructures}

\author[brk,lbl]{Jack Deslippe \corref{cor1}}
\ead{jdeslip@gmail.com}
%\affiliation{Department of Physics, University of California,
%Berkeley, California 94720} \affiliation{Materials Sciences
%Division, Lawrence Berkeley National Laboratory, Berkeley,
%California 94720}

\author[brk,lbl]{Georgy Samsonidze}
%\affiliation{Department of Physics, University of California,
%Berkeley, California 94720} \affiliation{Materials Sciences
%Division, Lawrence Berkeley National Laboratory, Berkeley,
%California 94720}

\author[brk,lbl]{David A. Strubbe}
%\affiliation{Department of Physics, University of California,
%Berkeley, California 94720} \affiliation{Materials Sciences
%Division, Lawrence Berkeley National Laboratory, Berkeley,
%California 94720}

\author[brk,lbl]{Manish Jain}
%\affiliation{Department of Physics, University of California,
%Berkeley, California 94720} \affiliation{Materials Sciences
%Division, Lawrence Berkeley National Laboratory, Berkeley,
%California 94720}

\author[brk,lbl]{Marvin L. Cohen}
%\affiliation{Department of Physics, University of California,
%Berkeley, California 94720} \affiliation{Materials Sciences
%Division, Lawrence Berkeley National Laboratory, Berkeley,
%California 94720}

\author[brk,lbl]{Steven G. Louie}
%\affiliation{Department of Physics, University of California,
%Berkeley, California 94720} \affiliation{Materials Sciences
%Division, Lawrence Berkeley National Laboratory, Berkeley,
%California 94720}

\cortext[cor1]{Corresponding author}
\address[brk]{Department of Physics, University of California,
Berkeley, California 94720}
\address[lbl]{Materials Sciences Division, Lawrence Berkeley
National Laboratory, Berkeley, California 94720}

\date{\today}

\begin{abstract}
\texttt{BerkeleyGW} is a massively parallel computational package for electron excited-state properties that is based on many-body perturbation theory employing the {\it ab initio} GW and GW plus Bethe-Salpeter equation methodology. It can be used in conjunction with many density-functional theory codes for ground-state properties, including \texttt{PARATEC}, \texttt{PARSEC}, \texttt{Quantum ESPRESSO}, \texttt{SIESTA}, and \texttt{Octopus}. The package can be used to compute the electronic and optical properties of a wide variety of material systems from bulk semiconductors and metals to nanostructured materials and molecules. The package scales to 10000s of CPUs and can be used to study systems containing up to 100s of atoms.
\end{abstract}


%\pacs{73.22.-f, 72.80.Rj, 75.70.Ak}

\maketitle

%%%%%%%%%%%%%%%%%%%%%%%%%%%%%%%%%%%%%%%%%%%%%%%%%%%%%%%%%%%%%%%%%%%
% The body of text starts.
%%%%%%%%%%%%%%%%%%%%%%%%%%%%%%%%%%%%%%%%%%%%%%%%%%%%%%%%%%%%%%%%%%%

\section{Program Summary}

\begin{figure} 
\begin{center} 
\includegraphics[width=10.0cm] {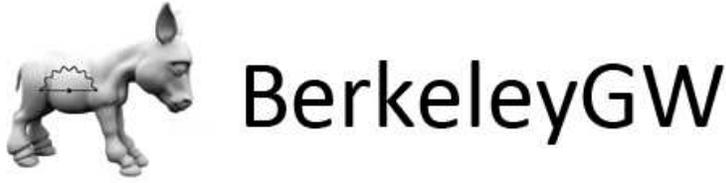} 
\end{center} 
\caption{The logo for the BerkeleyGW code.} 
\end{figure} 

Program title: BerkeleyGW

Program obtainable from: http://www.berkeleygw.org

Licensing provisions: See code for licensing.

No. of lines in distributed program, including test data, etc.: 80,000

No. of bytes in distributed program, including test data, output, etc.: 200MB

Distribution format: tar

Programming language: Fortran 90, C, C++, Python, Perl, BASH

Libraries required: BLAS, LAPACK, FFTW, ScaLAPACK (optional), MPI (optional). All available under open-source licenses.

Memory required: (50-2000) MB per CPU (Highly dependent on system size)

Computers for which the program has been designed and others on which it has been operable: Linux/UNIX workstations or clusters

Operating systems under which the program has been tested: Tested on a variety of Linux distributions in parallel and serial as well as AIX and Mac OSX. 

Nature of problem: The excited state properties of materials involve the addition or subtraction of electrons as well the optical excitations of electron-hole pairs. The excited particles interact strongly with other electrons in a material system. This interaction affects the electronic energies, wavefunctions and lifetimes. It is well known that ground-state theories, such as standard methods based on density-functional theory, fail to correctly capture this physics.

Solution method: We construct and solve Dyson's equation for the quasiparticle energies and wavefunctions within the GW approximation for the electron self energy.  We additionally construct and solve the Bethe-Salpeter equation for the correlated electron-hole (exciton) wavefunctions and excitation energies. 

Restrictions: The material size is limited in practice by the computational resources available. Materials with up to 500 atoms per periodic cell can be studied on large HPCs.

Running time: 1-1000 minutes (depending greatly on system size and processor number)

\section{Introduction}

Over the last few decades, the {\it ab initio} GW methodology has been successfully applied to the study of the quasiparticle properties of a large range of material systems from traditional bulk semiconductors, insulators and metals to, more recently, nano-systems like polymers, nano-wires and molecules \cite{hybertsen86,louiechapter06,spataru04,spataru04long,deslippe07}. The GW approach, which is based on approximating the electron self energy as the first term in an expansion in the screened Coulomb interaction, $W$ \cite{hedin65}, has proven to yield quantitatively accurate quasiparticle band gaps and dispersion relations from first principles. 

Additionally, the Bethe-Salpeter equation (BSE) approach to the optical properties of materials has proven exceptionally accurate in predicting the optical response of a similarly large class of materials employing an electron-hole interaction kernel derived within the same level of approximations as GW \cite{strinati88,rohlfing00,reining98,benedict98}.

The combined GW-BSE approach is now arguably regarded as the most accurate methodology commonly used for computing the quasiparticle and optical properties of condensed-matter systems. A perceived drawback of the GW methodology is its computational cost; a GW-BSE calculation is usually thought to be an order of magnitude (or worse) more costly than a typical density functional theory (DFT) calculation for the same system. Since the pioneering work of Ref. \cite{hybertsen86}, many GW implementations have been made, but most are limited to small systems of the size of 10s of atoms, and scaling to only small numbers of CPUs on the order of 100.

\texttt{BerkeleyGW} is a massively parallel computer package written predominantly in \texttt{FORTRAN90} that implements the {\it ab initio} GW methodology of Hybertsen and Louie \cite{hybertsen86} and includes many more recent advances, such as the Bethe-Salpeter equation approach for optical properties \cite{rohlfing00}. It alleviates the restriction to small numbers of atoms and scales beyond thousands of CPUs. The package is intended to be used on top of a number of mean-field (DFT and other) codes that focus on ground-state properties, such as \texttt{PARATEC} \cite{paratec}, \texttt{Quantum ESPRESSO} \cite{espresso}, \texttt{SIESTA} \cite{soler02siesta}, \texttt{PARSEC} \cite{parsec,parsec2}, \texttt{Octopus} \cite{octopus_pssb, octopus_CPC} and an empirical pseudopotential code (EPM) included in the package (based on TBPW \cite{martin04}). More information about \texttt{BerkeleyGW}, the latest source code, and help forums can be found by visiting the website at \texttt{http://berkeleygw.org/}.

%====================================================================================================
%====================================================================================================

\section{Theoretical Framework}

The {\it ab initio} GW-BSE approach is a many-body Green's-function methodology in which the only input parameters are the constituent atoms and the approximate structure of the system \cite{hybertsen86,rohlfing00}. Typical calculations of the ground- and excited-state properties using the GW-BSE method can be broken into three steps: (1) the solution of the ground-state structural and electronic properties within a suitable ground-state theory such as {\it ab initio} pseudopotential density-functional theory, (2) the calculation of the quasiparticle energies and wavefunctions within the GW approximation for the electron self-energy operator, and (3) the calculation of the two-particle correlated electron-hole excited states through the solution of a Bethe-Salpeter equation.

DFT calculations, often the chosen starting point for GW, are performed by solving the self-consistent Kohn-Sham equations with an approximate functional for the exchange-correlation potential, $V_{\rm xc}$ -- common approximations being the local density approximation (LDA) \cite{kohn65} and the generalized-gradient approximation (GGA) \cite{perdew96}:
\begin{equation}
\left[ -\frac{1}{2}\nabla^2+V_{\rm ion}+V_{\rm H}+V_{\rm xc}^{\rm DFT} \right] \psi_{n{\bf k}}^{\rm DFT}=E_{n{\bf k}}^{\rm DFT}\psi_{n{\bf k}}^{\rm DFT}
\end{equation}
where $E_{n\bf k}^{\text{DFT}}$ and $\psi_{n\bf k}^{\text{DFT}}$ are the Kohn-Sham eigenvalues and eigenfunctions respectively, $V_{\rm ion}$ is the ionic potential, $V_{\rm H}$ is the Hartree potential and $V_{\rm xc}$ is the exchange-correlation potential within a suitable approximation. When DFT is chosen as the starting point for GW, the Kohn-Sham wavefunctions and eigenvalues are used here as a first guess for their quasiparticle counterparts. The quasiparticle energies and wavefunctions (\textit{i.e.}, the one-particle excitations) are computed by solving the following Dyson equation \cite{hedin70,hybertsen86} in atomic units:
\begin{equation}
\label{dyson_equation}
\left[ -\frac{1}{2}\nabla^2+V_{\rm ion}+V_{\rm H}+\Sigma(E_{n\bf k}^{\rm QP}) \right] \psi_{n\bf k}^{\text{QP}}=E_{n\bf k}^{\rm QP}\psi_{n\bf k}^{\text{QP}}
\end{equation}
where $\Sigma$ is the self-energy operator within the GW approximation, and $E_{n\bf k}^{\rm QP}$ and $\psi_{n\bf k}^{\rm QP}$ are the quasiparticle energies and wavefunctions, respectively. For systems of periodic dimension less than three, the Coulomb interaction may be replaced by a truncated interaction. The interaction is set to zero for particle separation beyond the size of the system in order to avoid unphysical interaction between the material and its periodic images in a super-cell \cite{cohen75} calculation. The electron-hole excitation states (probed in optical or other measurements) are calculated through the solution of a Bethe-Salpeter equation \cite{rohlfing00,strinati88} for each exciton state $S$:
\begin{equation}
\bigl(E_{c\bf k}^{\rm QP}-E_{v\bf k}^{\rm QP}\bigr)A^S_{vc\bf k}+\sum_{v'c'\bf k'}\left<vc{\bf k}|K^{\rm eh}|v'c'{\bf k}'\right>
={\it\Omega}^SA^S_{vc\bf k}
\label{BSE}
\end{equation}
where $A^S_{vc\bf k}$ is the exciton wavefunction (in the quasiparticle state representation), ${\it\Omega}^S$ is the excitation energy, and $K^{\rm eh}$ is the electron-hole interaction kernel. We make the Tamm-Dancoff approximation by including only valence $\rightarrow$ conduction transitions \cite{rohlfing00,fetter}. The exciton wavefunction can be expressed in real space as:
\begin{equation}
\Psi({\bf r}_e,{\bf r}_h)=\sum_{{\bf k},c,v} A^S_{vc\bf k} \psi_{{\bf k},c}({\bf r}_e) \psi^{*}_{{\bf k},v}({\bf r}_h),
\label{plotxct}
\end{equation}
and the imaginary part of the dielectric function, if one is interested in optical properties, can be expressed as 
\begin{equation}
\epsilon_2(\omega)=\frac{16\pi^2e^2}{\omega^2}\sum_{S}\,\bigl|{\bf e}\cdot\left<0|{\bf v}|S\right>\bigr|^2
\delta\bigl(\omega-{\it\Omega}^S\bigr)
\label{eps2}
\end{equation}
where ${\bf e} \cdot \left<0|{\bf v}|S\right>$ is the velocity matrix element along the direction of the polarization of light, ${\bf e}$. One may compare this to the non-interacting absorption spectrum:
\begin{equation}
\epsilon_2(\omega)=\frac{16\pi^2e^2}{\omega^2}\sum_{vc\bf k}\,\bigl|{\bf e}\cdot\left<{v\bf k}|{\bf v}|{c\bf k}\right>\bigr|^2
\delta\bigl(\omega-E_{c\bf k}^{\rm QP}+E_{v\bf k}^{\rm QP}\bigr).
\end{equation}

\begin{figure}
\begin{center}
\includegraphics[width=6.324cm] {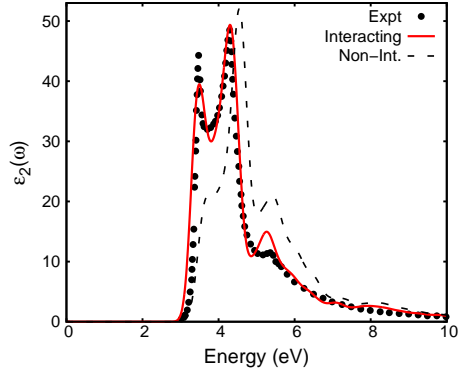}
\end{center}
\caption{The absorption spectra for silicon calculated at the GW (black dashed) and GW-BSE (red solid) levels using the \texttt{BerkeleyGW} package. Experimental data from \cite{jellison}.}
\label{siliconfig}
\end{figure}

An example absorption spectrum for silicon computed with the \texttt{BerkeleyGW} package at the GW and GW-BSE levels is shown in Fig. \ref{siliconfig}. Only when both the quasiparticle effects within the GW approximation and the excitonic effects through the solution of the Bethe-Salpeter equation are included is good agreement with experiment reached.

%====================================================================================================
%====================================================================================================

\section{Computational Layout}

\subsection{Major Sections of the Code}

\begin{figure}
\begin{center}
\includegraphics[width=8.2cm] {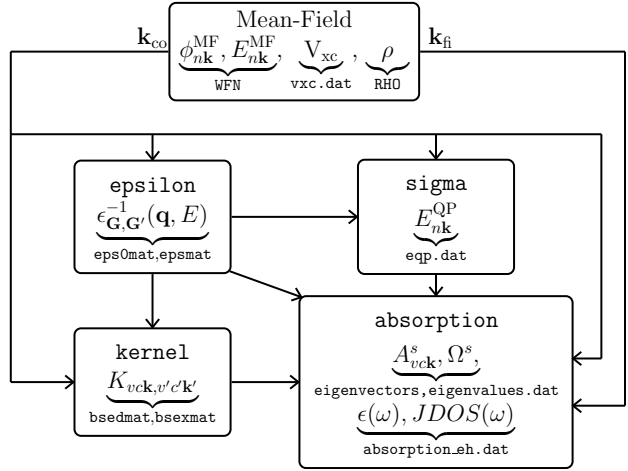}
\end{center}
\caption{Flow chart of a GW-BSE calculation performed in the \texttt{BerkeleyGW} package.}
\label{flowchart}
\end{figure}

Figure \ref{flowchart} illustrates the procedure for carrying out an {\it ab initio} GW-BSE calculation to obtain quasiparticle and optical properties using the \texttt{BerkeleyGW} code. First, one obtains the mean-field electronic orbitals and eigenvalues as well as the charge density. One can utilize one of the many supported DFT codes \cite{paratec,espresso,soler02siesta,parsec2,octopus_CPC} to construct this mean-field starting point and convert it to the plane-wave \texttt{BerkeleyGW} format (see Appendix) using the wrappers included. (Note that norm-conserving pseudopotentials must be used, or else extra contributions would need to be added to our matrix elements.)

The \texttt{Epsilon} executable produces the polarizability and inverse dielectric matrices. In the \texttt{epsilon} executable, the static or frequency-dependent polarizability and dielectric function are calculated within the random-phase approximation (RPA) using the electronic eigenvalues and eigenfunctions from a mean-field reference system. The main outputs are files \texttt{eps0mat} and \texttt{epsmat} that contain the inverse-dielectric matrix for ${\bf q} \rightarrow 0$ and ${\bf q} \ne 0$.

In the \texttt{sigma} executable, the screened Coulomb interaction, $W$, is constructed from the inverse dielectric matrix and the one-particle Green's function, $G$, is constructed from the mean-field eigenvalues and eigenfunctions. We then calculate the diagonal and (optionally) off-diagonal elements of the self-energy operator, $\Sigma=iGW$, as a matrix in the mean-field basis. In many cases, only the diagonal elements are sizable within the chosen mean-field orbital basis; in such cases, in applications to real materials, the effects of $\Sigma$ can be treated within first-order perturbation theory. The \texttt{sigma} executable evaluates $\Sigma$ in the form $\Sigma = V_{\rm xc} + (\Sigma - V_{\rm xc})$, where $V_{\rm xc}$ is the independent-particle mean-field approximation to the exchange-correlation potential of the chosen mean-field system. For moderately correlated electron systems, the best available mean-field Hamiltonian may often be taken to be the Kohn-Sham Hamiltonian \cite{kohn65}. However, many mean-field starting points are consistent with the \texttt{BerkeleyGW} package, such as Hartree-Fock, static COHSEX and hybrid functionals. In principle, the process of correcting the eigenfunctions and eigenvalues (which determine $W$ and $G$) could be repeated until self-consistency is reached or the $\Sigma$ matrix diagonalized in full. However, in practice, it is found that an adequate solution often is obtained within first-order perturbation theory on Dyson's equation for a given $\Sigma$ \cite{Holm98,Aryasetiawan98}.  Comparison of calculated energies with experiment shows that this level of approximation is very accurate for semiconductors and insulators and for most conventional metals. The outputs of the \texttt{sigma} executable are $E^{\rm QP}$, the quasiparticle energies, which are written to the file \texttt{eqp.dat} using the \texttt{eqp.py} post-processing utility on the generated \texttt{sigma.log} files for each \texttt{sigma} run.

The BSE executable, \texttt{kernel}, takes as input the full dielectric matrix calculated in the \texttt{epsilon} executable, which is used to screen the attractive direct electron-hole interaction, and the quasiparticle wavefunctions, which often are taken to be the same as the mean-field wavefunctions. The direct and exchange part of the electron-hole kernel are calculated and output into the \texttt{bsedmat} and \texttt{bsexmat} files respectively. The \texttt{absorption} executable uses these matrices, the quasiparticle energies and wavefunctions from a coarse $\textbf{k}$-point grid GW calculation, as well as the wavefunctions from a fine $\textbf{k}$-point grid. The quasiparticle energy corrections and the kernel matrix elements are interpolated onto the fine grid. The Bethe-Salpeter Hamiltonian, consisting of the electron-hole kernel with the addition of the kinetic-energy term, is constructed in the quasiparticle electron-hole pair basis and diagonalized yielding the electron-hole amplitude, or exciton wavefunctions, and excitation energies, printed in the file \texttt{eigenvectors}. Exciton binding energies can be inferred from the energy of the correlated exciton states relative to the inter-band-transition continuum edge. With the excitation energies and amplitudes of the electron-hole pairs, one then can calculate the macroscopic dielectric function for various light polarizations which is written to the file \texttt{absorption\_eh.dat}. This may be compared to the absorption spectrum without the electron-hole interaction included, printed in the file \texttt{absorption\_noeh.dat}.

Example input files for each executable are contained within the source code for the package, as well as complete example calculations for silicon, the (8,0) and (5,5) single-walled carbon nanotubes (SWCNTs), the CO molecule, and sodium metal. There are several post-processing and visualization utilities included in the package that are described in Sec. \ref{sec:utilities}.

Additionally, sums over ${\bf k}$ and ${\bf q}$ are accompanied
by an implicit division by the volume of the super-cell considered, $V_{\text{sc}} = N_k V_{\text{uc}}$, where $N_k$ is the number of points in the ${\bf k}$-grid and $V_{\text{uc}}$ is the volume of the unit cell in a periodic system.

Throughout the paper, we refer to benchmark numbers from calculations on the (20,20) SWCNT. This system has 80 carbon atoms and 160 occupied bands. We use 800 unoccupied bands in all sums requiring empty orbitals. We use a super-cell of size $80 \times 80 \times 4.6$ ${\rm au}^3$ equivalent to a bulk system of greater than 500 atoms. We use a $1 \times 1 \times 32$ coarse {\bf k}-grid and a $1 \times 1 \times 256$ fine {\bf k}-grid. We calculate the self-energy corrections within the diagonal approximation for 8 conduction and 8 valence bands. The Bethe-Salpeter equation is solved with 8 conduction and 8 valence bands. The relative costs of the various steps in the GW-BSE calculation using the \texttt{BerkeleyGW} package is shown in Table \ref{timeBreakdown}. As can be seen from the table, the actual time to solution for the GW-BSE part of the calculation is smaller than that of the DFT parts.

\begin{figure}
\begin{center}
\includegraphics[width=5.2cm] {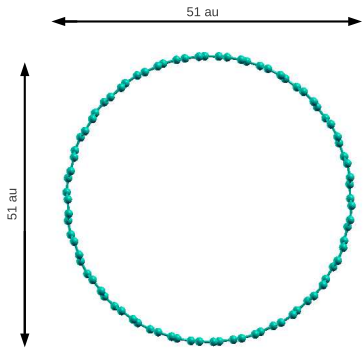}
\end{center}
\caption{The cross section of the (20,20) SWCNT used throughout the paper as a benchmark system.}
\label{20-20-structure}
\end{figure}

\begin{table}
\begin{center}
\begin{tabular}{ | c | c | c | c | }  \hline 
Step & \# CPUs & CPU hours & Wall hours \\ \hline
DFT Coarse & 64 $\times$ 32 & 19000 & 9.1 	\\ 
DFT Fine & 64 $\times$ 256 & 29000 & 1.8 	\\ 
\texttt{epsilon} & 1600 $\times$ 32 & 61000 & 1.2 \\
\texttt{sigma} & 960 $\times$ 16 & 46000 & 3.0 \\
\texttt{kernel} & 1024 & 600 & 0.6 \\
\texttt{absorption} & 256 & 500 & 2.0 \\
\hline
\end{tabular}
\end{center}
\caption{Breakdown of the CPU and wall-clock time spent on the calculation of the (20,20) SWCNT with parameters described in the text. The $\times$ indicates an additional level of trivial parallelization over the {\bf k}- or {\bf q}-points.}
\label{timeBreakdown}
\end{table}

%================================================================================
%================================================================================

\subsection{RPA Dielectric Matrix: \texttt{epsilon}}
\label{sec:xi0}

\texttt{epsilon} is a standalone executable that computes either the static or dynamic RPA polarizability and corresponding inverse dielectric function from input electronic eigenvalues and eigenvectors computed in a suitable mean-field code. As we discuss in detail below, the input electronic eigenvalues and eigenvectors can come from a variety of different mean-field approximations including DFT within LDA/GGA, generalized Kohn-Sham hybrid-functional approximations as well as direct approximations to the GW Dyson's equation such as the static-COHSEX \cite{hedin70,mjain10COHSEX} approximation and the Hartree-Fock approximation.

\begin{figure}
\begin{center}
\includegraphics[width=6.5cm] {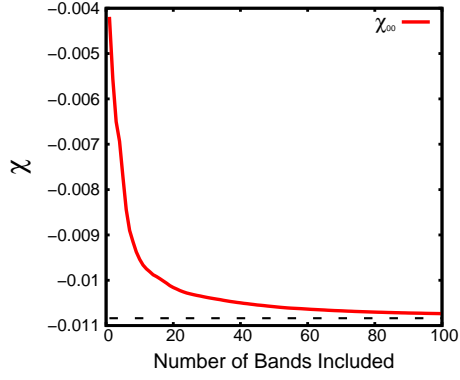}
\end{center}
\caption{Example convergence output plotted from \texttt{chi\_converge.dat} showing the convergence of the sum in Eq. \ref{eqn_static_xi} for the ${\bf G},{\bf G}'=0$ and ${\bf q}=(0,0,0.5)$ component of $\chi$ in ZnO.}
\label{xi0Converge}
\end{figure}

We will first discuss the computation of the static polarizability and the inverse dielectric matrix. The \texttt{epsilon} executable computes the static RPA polarizability using the following expression \cite{hybertsen86}:
%\begin{widetext}
\begin{equation}
\chi_{{\bf GG}'}{\left({\bf q}\;\!;0\right)}=
\,\,{}\sum_{n}^{\rm occ}\sum_{n'}^{\rm emp}\sum_{{\bf k}}
M_{nn'}^{*}({\bf k},{\bf q},{\bf G})
M_{nn'}({\bf k},{\bf q},{\bf G'})
\frac{1}{E_{n{\bf k}{+}{\bf q}}\,{-}\,E_{n'{\bf k}}}.
\label{eqn_static_xi}
\end{equation}
%\end{widetext}
where
\begin{equation}
M_{nn'}({\bf k},{\bf q},{\bf G})=\left<n{\bf k}{+}{\bf q}\right|e^{i({\bf q}+{\bf G})\cdot{\bf r}}\left|n'{\bf k}\right>
\label{eqn_matrix_elem}
\end{equation}
are the plane-wave matrix elements. Here ${\bf q}$ is a vector in the first Brillouin zone, ${\bf G}$ is a reciprocal-lattice vector, and $\left<n{\bf k}\right| $ and $E_{n{\bf k}}$ are the mean-field electronic eigenvectors and eigenvalues. The matrix in Eq. \ref{eqn_static_xi} is to be evaluated up to $|{\bf q} + {\bf G}|^2,|{\bf q} + {\bf G}'|^2 < E_{\text{cut}}$ where $E_{\text{cut}}$ defines the dielectric energy cutoff. The number of empty states, $n'$, included in the summation must be such that the highest empty state included has an energy corresponding to $E_{\text{cut}}$. There are therefore not two convergence parameters, but only one, in evaluating Eq. \ref{eqn_static_xi}: one either must choose to converge with empty states or with the dielectric energy cutoff and set the remaining parameter to match the chosen convergence parameter. The \texttt{epsilon} code itself reports the convergence of Eq. \ref{eqn_static_xi} in an output file called \texttt{chi\_converge.dat} (plotted in Fig. \ref{xi0Converge}), that presents the computed value of $\chi_{{\bf GG}'=0}{\left({\bf q}\;\!;0\right)}$ and $\chi_{{\bf GG}'={\bf G}_{\rm max}}{\left({\bf q}\;\!;0\right)}$ using partial sums in Eq. \ref{eqn_static_xi} where ${\bf G}_{\rm max}$ is the largest reciprocal-lattice vector included, and the number of empty states is varied between 1 and the maximum number requested in the input file, \texttt{epsilon.inp}. A simple extrapolation also is included.

With the expression for $\chi$ above, we can obtain the RPA dielectric matrix as
\begin{equation}
\epsilon_{{\bf GG}'}{\left({\bf q}\;\!;0\right)}=
\delta_{{\bf GG}'}\,{-}\,v{\left({\bf q}{+}{\bf G}\right)}
\chi_{{\bf GG}'}{\left({\bf q}\;\!;0\right)}
\label{epsilon}
\end{equation}
where $v{\left({\bf q}{+}{\bf G}\right)}$ is the bare Coulomb interaction defined as:
\begin{equation}
v{\left({\bf q}{+}{\bf G}\right)}=\frac{4 \pi}{\left|{\bf q} + {\bf G}\right|^2}
\label{untruncated_bare_coulomb}
\end{equation}
in the case of bulk crystals where no truncation is necessary. We discuss in Sec. \ref{sec:truncation} how to generalize this expression for the case of nano-systems where truncating the interaction in non-periodic directions greatly improves the convergence with super-cell size.

It should be noted that we use an asymmetric definition of the Coulomb interaction, as opposed to symmetric expressions such as
\begin{align}
v({\bf q} + {\bf G},{\bf q} + {\bf G'}) = \frac{4 \pi}{\left| {\bf q} + {\bf G} \right| \left| {\bf q} + {\bf G'} \right| }.
\end{align}
This causes $\epsilon_{{\bf GG}'}{\left({\bf q}\;\!;0\right)}$ and $\chi_{{\bf GG}'}{\left({\bf q}\;\!;0\right)}$ to be also asymmetric in ${\bf G}$ and ${\bf G}'$. This asymmetry is resolved when constructing the static screened Coulomb interaction by use of the expression:
\begin{equation}
W_{{\bf GG}'}{\left({\bf q}\;\!;0\right)}=\epsilon^{-1}_{{\bf GG}'}{\left({\bf q}\;\!;0\right)}v{\left({\bf q}{+}{\bf G}'\right)}.
\label{screened_interaction}
\end{equation}
Here $W$ is symmetric in ${\bf G}$ and ${\bf G}'$ even though both $v$ and $\epsilon^{-1}$ individually are not.

The computation of $\epsilon^{-1}_{{\bf GG}'}{\left({\bf q}\;\!;0\right)}$ in the \texttt{epsilon} code involves three computationally intensive steps: the computation of the matrix elements needed for the summation in Eq. \ref{eqn_static_xi}, the summation itself and the inversion of the dielectric matrix to yield $\epsilon^{-1}_{{\bf GG}'}{\left({\bf q}\;\!;0\right)}$.
The \texttt{epsilon} code first computes all the matrix elements $M_{nn'}$ required in the summation for Eq. \ref{eqn_static_xi}. This step is generally the most time-consuming step in the execution of the \texttt{epsilon} code. Naively, one might think this process scales as $N^4$, where $N$ is the number of atoms in the system. This is because both the number of valence and conduction bands needed scales linearly with $N$ and the number of ${\bf G}$ vectors scales linearly with the cell volume which itself scales linearly with the number of atoms.  Thus, we must calculate $N^3$ matrix elements each of which involves a sum over the plane-wave basis set for the eigenfunctions. We therefore have an $N^4$ scaling. However, we can achieve $N^3 \log N$ scaling by using fast Fourier transforms (FFTs), noting that the expression in Eq. \ref{eqn_matrix_elem} is a convolution in Fourier space \cite{fleszarthesis}.  Therefore, Eq. \ref{eqn_matrix_elem} can be written as the Fourier transform of a direct product of the wavefunctions in real space:
\begin{equation}
M_{nn'}({\bf k},{\bf q},\{{\bf G}\}) = {\rm FFT}^{-1}\left( \phi^{*}_{n,{\bf k}+{\bf q} }({\bf r}) \phi_{n',{\bf k} }({\bf r}) \right).
\end{equation}
The FFTs are implemented with FFTW \cite{fftw} and scale as $N \log N$. The computation of all the matrix elements needed for Eq. \ref{eqn_static_xi} therefore scales as $N^3 \log N$. We discuss in the following sections that the computation of these matrix elements can be parallelized very trivially up to tens of thousands of CPUs. Given an infinite resource of CPUs, our implementation would have a wall-time scaling of $N \log N$, nearly linear in the number of atoms. 

Having computed the individual matrix elements required in Eq. \ref{eqn_static_xi}, we now turn our attention to the summation involved in the same expression. It should be noted that the formal scaling of this step with the number of atoms is $N^4$ since one must sum over the number of occupied bands and the number of unoccupied bands for every ${\bf G}$ and ${\bf G'}$ pair -- each one of these quantities scales linearly with the number of atoms. This step therefore formally has the worst scaling of the entire GW process -- leading many to claim that GW as a whole scales like $N^4$. However, in practice for most systems currently under study within a generalized plasmon-pole (GPP) \cite{hybertsen86} or other approximation where this sum is done only once for the static polarizability, this step represents less than 10 percent of a typical calculation time even for systems of 100s of atoms because this step can be optimized and parallelized greatly. In particular, Eq. \ref{eqn_static_xi} can be written very compactly as a single matrix-matrix product for each ${\bf q}$:
\begin{equation}
\chi_{{\bf GG}'}{\left({\bf q}\;\!;0\right)}=
\,\,{\bf M}^{*}({\bf G},{\bf q},(n,n',{\bf k}))\cdot {\bf M}^{\rm T}({\bf G'},{\bf q},(n,n',{\bf k}))
\label{eqn_static_xi_compact}
\end{equation}
where $\left(n,n',{\bf k} \right)$ represents a single composite index that is summed over as the inner dimension in the matrix-matrix product. The matrices ${\bf M}$ can be expressed in terms of the matrix elements $M$ as:
\begin{equation}
{\bf M}({\bf G},{\bf q},(n,n',{\bf k})) = M_{nn'}({\bf k},{\bf q},{\bf G}) \cdot \frac{1}{\sqrt{E_{n{\bf k}{+}{\bf q}}\,{-}\,E_{n'{\bf k}}}}.
\end{equation}

The single dense matrix-matrix product required in Eq. \ref{eqn_static_xi_compact} still scales as $N^4$ since the inner dimension, $\left( n,n',{\bf k} \right)$, scales as $N^2$ and dense matrix multiplication itself scales as $N^2$.  However, in the \texttt{BerkeleyGW} package, this single step is still made quite rapid for even systems as large as 100s of atoms.   The LEVEL 3 BLAS \cite{blas} libraries \texttt{DGEMM} and \texttt{ZGEMM} and their parallel analogues are used to compute the single matrix product in Eq. \ref{eqn_static_xi_compact}.  As we discuss further in Sec. \ref{sec:xi0_parallel}, in the evaluation of Eq. \ref{eqn_matrix_elem}, the parallel wall-time scaling is $N^2$ with the number of atoms.

Finally, once we have constructed $\chi_{{\bf GG}'}{\left({\bf q}\;\!;0\right)}$ we can construct the RPA dielectric matrix and inverse dielectric matrix required for the computation of the screened Coulomb interaction, $W$. The dielectric matrix as implemented in the code is expressed in Eq. \ref{epsilon}.

Here we require for the first time the Coulomb interaction in reciprocal space $v{\left({\bf q}{+}{\bf G}\right)}$, which can be computed trivially from Eq. \ref{untruncated_bare_coulomb} for the case of bulk crystals, but requires an FFT for the case of nanostructured materials. We discuss this more in Sec. \ref{sec:truncation}.  

There is a clear problem in directly computing $\epsilon_{{\bf 00}}{\left({\bf q}={\bf 0}\right)}$ due to the fact that the Coulomb interaction, Eq. \ref{untruncated_bare_coulomb}, diverges as ${\bf q}\rightarrow 0$ except in the case of box-type truncation schemes (see Sec. \ref{sec:truncation}). For semiconducting systems, due to orthogonality, the matrix elements (Eq. \ref{eqn_matrix_elem}) themselves go to 0 with the form $|M_{nn'}({\bf k},{\bf q},{\bf G=0})| \propto |q|$. Thus $\epsilon({\bf q}\rightarrow0)$ contains a non-trivial ${q^2}/{q^2}$ limit. One way to handle this would be to take the limit of Eqs. \ref{eqn_static_xi} and \ref{eqn_matrix_elem} analytically via ${\bf k} \cdot {\bf p}$ perturbation theory, where the perturbation is the momentum operator $-i \nabla$ plus the commutators with the non-local potential of the mean-field Hamiltonian \cite{baldereschi78, hybertsen86}. This is analogous to the treatment of the velocity operator in \texttt{absorption} (Eq. \ref{velocityop}).

The \texttt{epsilon} code has implemented a simpler scheme, however, in which we numerically take the limit as ${\bf q}\rightarrow 0$ by evaluating $\epsilon_{{\bf 00}}{\left({\bf q}_{0}\right)}$ at a small but finite ${\bf q}_0$ usually taken as approximately 1/1000th of the Brillouin zone, in one of the periodic directions.  For semiconducting systems, where $\epsilon_{{\bf 00}}{\left({\bf q}={\bf 0}\right)}\rightarrow C$, it is sufficient to construct a separate ${\bf k}$-grid for the conduction and valence bands shifted by the small vector ${\bf q}_0$ in order to compute $M_{nn'}({\bf k},{\bf q}_{0},{\bf G=0})$, where $n$ is a valence and $n'$ a conduction band, and to evaluate the correct limiting ${q^2}/{q^2}$ ratio. For metals, however, intra-band transitions have $|M_{nn'}({\bf k},{\bf q},{\bf G=0})| \propto C$, yielding $\epsilon_{{\bf 00}}{\left({\bf q}\rightarrow0\right)}\propto {C'}/{q^2}$. In this case, the two-${\bf k}$-grid treatment is insufficient, because the proportionality coefficient $C'$ depends sensitively on the density of states (DOS) at the Fermi energy. Therefore a $\textbf{k}$-grid sampling of the same spacing as ${\bf q}_0$ is required, although fewer conduction bands are necessary in the sum since $\epsilon({\bf q}\rightarrow 0)$ is dominated by intra-band transitions. Thus we typically calculate $\epsilon({\bf q}\rightarrow 0)$ using a single fine wavefunction grid by using the smallest ${\bf q}$ consistent with the grid. Note that this treatment of intra-band transitions is still the zero-temperature limit in our code, as the effect of thermal occupations is small in GW except at very large temperatures \cite{hightemp}. Effectively occupations are taken as one below the Fermi level, zero above the Fermi level, and $1/2$ at the Fermi level (as needed for graphene at the Dirac point). This is despite any smearing that may have been used in the underlying mean-field calculation. We should point out that only one ${\bf q}_0$ is used; if the material is anisotropic (in periodic directions), in principle an average over the three directions of ${\bf q}_0$ should be done. This may be accomplished by using a vector in the (111) direction (referred to the principal axes of $\epsilon$). Neglect of the anisotropy can give significant errors in \texttt{sigma} \cite{sahar}.

The inversion of the dielectric matrix required to compute $W$, Eq. \ref{screened_interaction}, is done with LAPACK and ScaLAPACK (for parallel calculations) using \texttt{ZGESV}, \texttt{DGESV} and their parallel counterparts. The inversion scales like $N^3$ with the number of atoms and, as we discuss below, scales well up to 100s of processors with ScaLAPACK. In general, for systems of up to 100s of atoms, the inversion step represents less than 10 percent of the total computation time for \texttt{epsilon}.

We have so far limited ourselves to situations in which only a direct calculation of the static polarizability, Eq. \ref{eqn_static_xi}, is required, such as in the static-COHSEX approximation \cite{mjain10COHSEX} or when utilizing a GPP model \cite{hybertsen86} to extend the dielectric response to non-zero frequencies. However, we can also do a more refined calculation. Options are given in the code so that the dielectric matrix is computed directly at real frequencies without extrapolation, as is formally required in the Dyson equation. We use in the package the advanced and retarded dielectric functions, defined as:

%\begin{widetext}
\begin{eqnarray}
\label{retarded_advanced_epsilon}
\epsilon_{{\bf GG}'}^{\rm r/a}{\left({\bf q}\;\!;E\right)} &=&
\delta_{{\bf GG}'}-v{\left({\bf q}{+}{\bf G}\right)} \\ \nonumber
&\times& \sum_{n}^{\rm occ}\sum_{n'}^{\rm emp}\sum_{{\bf k}}
M_{nn'}^{*}({\bf k},{\bf q},{\bf G})
M_{nn'}({\bf k},{\bf q},{\bf G'}) \\ \nonumber
%\left<n{\bf k}{+}{\bf q}\right|e^{i({\bf q}+{\bf G})\cdot{\bf r}}\left|n'{\bf k}\right>
%\left<n'{\bf k}\right|e^{-i({\bf q}+{\bf G}')\cdot{\bf r}'}\left|n{\bf k}{+}{\bf q}\right> \\
&\times& \frac{1}{2}
\left[\,\frac{1}{E_{n{\bf k}{+}{\bf q}}\,{-}\,E_{n'{\bf k}}\,{-}\,E\,{\mp}\,i\delta}
\,{+}\,\frac{1}{E_{n{\bf k}{+}{\bf q}}\,{-}\,E_{n'{\bf k}}\,{+}\,E\,{\pm}\,i\delta}\,\right]
\end{eqnarray}
%\end{widetext}
where $E$ is the evaluation frequency and $\delta$ is a broadening parameter chosen to be consistent with the energy spacing afforded by the $\textbf{k}$-point sampling of the calculation, using the upper (lower) signs for the retarded (advanced) function. In principle, one must converge the calculation with respect to increasing the $\textbf{k}$-point sampling and decreasing this broadening parameter.

\begin{figure}
\begin{center}
\includegraphics[width=8.5cm] {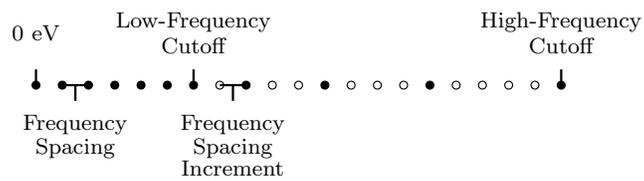}
\end{center}
\caption{Schematic of the frequency-grid parameters for a full-frequency calculation in \texttt{epsilon}. The open circles are a continuation of the uniform grid that are omitted above the low-frequency cutoff.}
\label{frequency_grid}
\end{figure}

In the \texttt{epsilon} code, we compute Eq. \ref{retarded_advanced_epsilon} on a grid of real frequencies, $E$, specified by a frequency spacing, a low-frequency cutoff, a high-frequency cutoff and a frequency-spacing increment. We sample the frequency on the real axis uniformly from 0 to the low-frequency cutoff with a sampling rate given by the frequency spacing.  We then increase the frequency spacing by the step increment until we reach the high-frequency cutoff (Fig. \ref{frequency_grid}). In general, one also must refine this frequency grid until convergence is reached, but we find that for the purpose of calculating band gaps of typical semiconductors, a frequency spacing of a few hundred meV and a high-frequency cutoff of twice the dielectric energy cutoff is sufficient, though it should be noted this energy can be quite high (\textit{e.g.} the case of ZnO \cite{shih10}).

\begin{figure}
\begin{center}
\includegraphics[width=6.5cm] {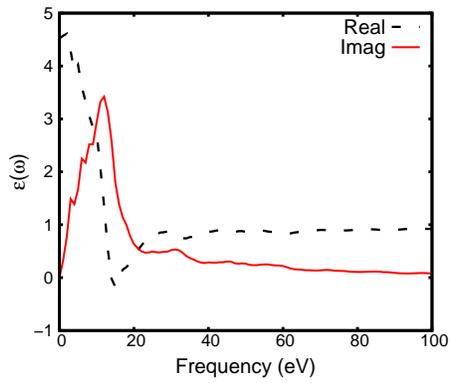}
\end{center}
\caption{Example output plotted from \texttt{EpsDyn} file showing the computed $\epsilon_{\bf 00}(\omega)$ in ZnO.}
\label{EpsDyn}
\end{figure}

In cases where the computation of the ``full-frequency'' dielectric response function is required, the bottleneck of the calculation often does become the $N^4$ summation step of Eq. \ref{eqn_static_xi_compact}.  This is because the computation of the matrix elements, Eq. \ref{eqn_matrix_elem}, needs only be done once, whereas the summation must be done for all frequencies separately. Because of this, a full-frequency \texttt{epsilon} calculation of between 10-50 frequencies costs only twice the time of a static \texttt{epsilon} calculation, but the cost scales linearly with frequencies after this point.

%=====================================================================================
%=====================================================================================

\subsection{Computation of the Self-Energy: \texttt{sigma}}
\label{sec:sigma}

The \texttt{sigma} executable takes as input the inverse epsilon matrix calculated from the \texttt{epsilon} executable and a suitable set of mean-field electronic energies and wavefunctions.  It computes a representation of the Dyson's equation, Eq. \ref{dyson_equation}, in the basis of the mean-field eigenfunctions through the computation of the diagonal and off-diagonal elements of $\Sigma$:
\begin{align}
\label{QP-offdiag}
\langle \psi_{n{\bf k}} | & H^{\rm QP}(E) | \psi_{m{\bf k}} \rangle = \\
\nonumber
& E^{\rm MF}_{n{\bf k}}\delta_{n,m} + \langle \psi_{n{\bf k}} | \Sigma \left( E \right) - \Sigma^{\rm MF} \left(E \right) | \psi_{m{\bf k}} \rangle
\end{align}
where $E$ is an energy parameter that should be set self-consistently to the quasiparticle eigenvalues, $E^{\rm MF}_{n{\bf k}}$ and $\psi_{n{\bf k}}$ are the mean-field eigenvalues and eigenvectors and $\Sigma^{\text{MF}}$ is a mean-field approximation to the electronic self-energy operator, such as $V_{\rm xc}$ in the case of a DFT starting point.

It is often the case that the mean-field wavefunctions are sufficiently close to the quasiparticle wavefunctions \cite{hybertsen86} that one may reduce Eq. \ref{QP-offdiag} to include only diagonal matrix elements. In this case the user may ask for only diagonal elements, and the quasiparticle energies will be updated in the following way:
\begin{equation}
\label{QP-diag}
E^{\rm QP}_{n{\bf k}} = E^{\rm MF}_{n{\bf k}} + \langle \psi_{n{\bf k}} | \Sigma \left( E \right) - \Sigma^{\rm MF} \left( E \right) | \psi_{n{\bf k}} \rangle.
\end{equation} 
The mean field in Eq. \ref{QP-diag} and Eq. \ref{QP-offdiag} can be DFT within the LDA or GGA schemes as well as within a hybrid-functional approach. In the LDA case, for example, $\Sigma^{\rm MF} \left( E \right) = V_{\rm xc}$, is local and energy-independent. The starting mean-field calculation can also be an approximation to the Dyson's equation, Eq. \ref{dyson_equation}, such as Hartree-Fock (the zero-screening limit) or static COHSEX (the static-screening limit) \cite{rinke05,bruneval06,mjain10COHSEX}. The use of these mean-field starting points for construction of Eq. \ref{QP-offdiag} and Eq. \ref{QP-diag} is classified as a one-shot $G_0W_0$ calculation (the $0$ subscript means that both $G$ and $W$ are constructed from the mean-field eigenvalues and eigenvectors). One also can start from a previous iteration of GW in an eigenvalue or eigenvector self-consistency scheme \cite{bruneval06,vanschilfgaarde06}. In this case, the `MF' superscripts in Eq. \ref{QP-diag} and \ref{QP-offdiag} should be renamed ``previous'' to designate the self-consistency process.

The \texttt{sigma} executable itself can evaluate the matrix elements of $\Sigma$ in Eq. \ref{QP-diag} and Eq. \ref{QP-offdiag} within various approximations: Hartree-Fock, static COHSEX, GW within a GPP model and full-frequency GW.  

For GW and static-COHSEX calculations, $\Sigma$ can be broken into two parts, $\Sigma = \Sigma_{\rm SX} + \Sigma_{\rm CH}$, where $\Sigma_{\rm SX}$ is the screened exchange operator and $\Sigma_{\rm CH}$ is the Coulomb-hole operator \cite{hybertsen86,hedin65,hedin70}. These are implemented in the \texttt{sigma} executable in the following way for a full-frequency calculation:
%\begin{widetext}
\begin{align}
\label{ff-sx}
\left<n{\bf k}\right|\Sigma_{\rm SX}{\left(E\right)}\left|n'{\bf k}\right>&=
-\sum_{n''}^{\rm occ}\sum_{{\bf qGG}'}
M^{*}_{n''n}({\bf k},-{\bf q},-{\bf G})M_{n''n'}({\bf k},-{\bf q},-{\bf G}') %\\
%\left<n{\bf k}\right|e^{i({\bf q}+{\bf G})\cdot{\bf r}}\left|n''{\bf k}{-}{\bf q}\right>
%\left<n''{\bf k}{-}{\bf q}\right|e^{-i({\bf q}+{\bf G}')\cdot{\bf r}'}\left|n'{\bf k}\right> \\
\\ \nonumber
&\times
\left[ \epsilon_{{\bf GG}'}\right]^{-1}{\left({\bf q}\;\!;E\,{-}\,E_{n''{\bf k}{-}{\bf q}}\right)}
v{\left({\bf q}{+}{\bf G}'\right)}
\end{align}
and 
\begin{align}
\label{ff-ch}
\left<n{\bf k}\right|\Sigma_{\rm CH}{\left(E\right)}\left|n'{\bf k}\right>&=
\frac{i}{2\pi}\sum_{n''}\sum_{{\bf qGG}'}
M^{*}_{n''n}({\bf k},-{\bf q},-{\bf G})M_{n''n'}({\bf k},-{\bf q},-{\bf G}') \\
%\left<n{\bf k}\right|e^{i({\bf q}+{\bf G})\cdot{\bf r}}\left|n''{\bf k}{-}{\bf q}\right>
%\left<n''{\bf k}{-}{\bf q}\right|e^{-i({\bf q}+{\bf G}')\cdot{\bf r}'}\left|n'{\bf k}\right> \\
\nonumber
&\times\,\int_0^\infty\!\!dE'\,\,
\frac{\left[ \epsilon_{{\bf GG}'}^{\rm r} \right]^{-1}{\left({\bf q}\;\!;E'\right)}
\,{-}\,\left[\epsilon_{{\bf GG}'}^{\rm a} \right]^{-1}{\left({\bf q}\;\!;E'\right)}}
{E\,{-}\,E_{n''{\bf k}{-}{\bf q}}\,{-}\,E'\,{+}\,i\delta}\,\,
\;v{\left({\bf q}{+}{\bf G}'\right)}
\end{align}
%\end{widetext}
where $M$ is defined in Eq. \ref{eqn_matrix_elem} and $\epsilon^{\rm r}$ and $\epsilon^{\rm a}$ are the retarded and advanced dielectric matrices defined in Eq. \ref{retarded_advanced_epsilon} \cite{spataruthesis}. In practice the \texttt{sigma} executable computes the matrix elements of bare exchange, $\Sigma_{\rm X}$ and of $\Sigma_{\rm SX}-\Sigma_{\rm X}$, where the matrix elements of $\Sigma_{\rm X}$ are obtained by replacing $\left[ \epsilon_{{\bf GG}'}\right]^{-1}{\left({\bf q}\;\!;E\,{-}\,E_{n''{\bf k}{-}{\bf q}}\right)}$ with $\delta_{{\bf G},{\bf G'}}$ in Eq. \ref{ff-sx} (as given by Eq. \ref{hf-x} below). The integral in Eq. \ref{ff-ch} over frequency is done numerically on the frequency grid used in the \texttt{epsilon} executable (Fig. \ref{frequency_grid}). 

For GPP calculations, the corresponding expressions used in the code are:
%\begin{widetext}
\begin{align}
\label{gpp-sx}
\left<n{\bf k}\right|\Sigma_{\rm SX}{\left(E\right)}\left|n'{\bf k}\right>=
-\sum_{n''}^{\rm occ}\sum_{{\bf qGG}'}
M^{*}_{n''n}({\bf k},-{\bf q},-{\bf G})M_{n''n'}({\bf k},-{\bf q},-{\bf G}') \\
%\left<n{\bf k}\right|e^{i({\bf q}+{\bf G})\cdot{\bf r}}\left|n''{\bf k}{-}{\bf q}\right>
%\left<n''{\bf k}{-}{\bf q}\right|e^{-i({\bf q}+{\bf G}')\cdot{\bf r}'}\left|n'{\bf k}\right>\\
\nonumber
\times
\left[\delta_{{\bf GG}'}+\frac{\Omega^2_{{\bf GG}'}{\left({\bf q}\;\!\right)}
\left(1\,{-}\,i\tan\phi_{{\bf GG}'}{\left({\bf q}\;\!\right)}\right)}
{\left(E\,{-}\,E_{n''{\bf k}{-}{\bf q}}\right)^2\!{-}\:
\tilde{\omega}^2_{{\bf GG}'}{\left({\bf q}\;\!\right)}}\right]
v{\left({\bf q}{+}{\bf G}'\right)}
\end{align}
and
\begin{align}
\label{gpp-ch}
\left<n{\bf k}\right|\Sigma_{\rm CH}{\left(E\right)}\left|n'{\bf k}\right>=
\frac{1}{2}\sum_{n''}\sum_{{\bf qGG}'}
M^{*}_{n''n}({\bf k},-{\bf q},-{\bf G})M_{n''n'}({\bf k},-{\bf q},-{\bf G}') \\
%\left<n{\bf k}\right|e^{i({\bf q}+{\bf G})\cdot{\bf r}}\left|n''{\bf k}{-}{\bf q}\right>
%\left<n''{\bf k}{-}{\bf q}\right|e^{-i({\bf q}+{\bf G}')\cdot{\bf r}'}\left|n'{\bf k}\right>\\
\nonumber
\times\,\frac{\Omega^2_{{\bf GG}'}{\left({\bf q}\;\!\right)}
\left(1\,{-}\,i\tan\phi_{{\bf GG}'}{\left({\bf q}\;\!\right)}\right)}
{\tilde{\omega}_{{\bf GG}'}{\left({\bf q}\;\!\right)}
\left(E\,{-}\,E_{n''{\bf k}{-}{\bf q}}{-}\,
\tilde{\omega}_{{\bf GG}'}{\left({\bf q}\;\!\right)}\right)}
\;v{\left({\bf q}{+}{\bf G}'\right)}
\end{align}
%\end{widetext}
where $\Omega_{{\bf GG}'}{\left({\bf q}\;\!\right)}$, $\tilde{\omega}_{{\bf GG}'}{\left({\bf q}\;\!\right)}$, $\lambda_{{\bf GG}'}{\left({\bf q}\;\!\right)}$ and $\phi_{{\bf GG}'}{\left({\bf q}\;\!\right)}$ are the effective bare plasma frequency, the GPP mode frequency, the amplitude and the phase of the renormalized $\Omega^2_{{\bf GG}'}{\left({\bf q}\;\!\right)}$ \cite{hybertsen86,zhang89} defined as:
\begin{equation}
\label{capitalomega}
\Omega^2_{{\bf GG}'}{\left({\bf q}\;\!\right)}=\omega_{\rm p}^2\,\,\frac{{\left({\bf q}{+}{\bf G}\right)}{\cdot}{\left({\bf q}{+}{\bf G}'\right)}}{\left|{\bf q}{+}{\bf G}\right|^2}\,\,\frac{\rho{\left({\bf G}{-}{\bf G}'\right)}}{\rho{\left({\bf 0}\right)}}
\end{equation}
\begin{equation}
\label{tildeomega}
\tilde{\omega}^2_{{\bf GG}'}{\left({\bf q}\;\!\right)}=\frac{\left| \lambda_{{\bf GG}'}{\left({\bf q}\;\!\right)} \right|} {\cos\phi_{{\bf GG}'}{\left({\bf q}\;\!\right)}}
\end{equation}
\begin{equation}
\label{phi}
\left| \lambda_{{\bf GG}'}{\left({\bf q}\;\!\right)} \right| e^{i\phi_{{\bf GG}'}{\left({\bf q}\;\!\right)}}=\frac{\Omega^2_{{\bf GG}'}{\left({\bf q}\;\!\right)}}{\delta_{{\bf G}{\bf G}'}{-}\epsilon_{{\bf G}{\bf G}'}^{-1}({\bf q};0)}
\end{equation}
Here, $\rho$ is the electron charge density in reciprocal space and $\omega_{\rm p}^2=4 \pi \rho({\bf 0}) e^2 / m$ is the classical plasma frequency. In this case, the integral over energy that is necessary in the full-frequency expression, Eq. \ref{ff-ch}, is reduced to a single term using an analytical approximation to the frequency dependence of the dielectric matrix requiring only the static dielectric matrix $\epsilon_{{\bf G}{\bf G}'}^{-1}({\bf q};0)$ in Eq. \ref{tildeomega}. The analytical approximation is done using the $f$-sum rule for each ${\bf G}{\bf G}'$ pair as described in Ref. \cite{hybertsen86}. This reduces the computational cost of evaluating the $\Sigma$ matrix elements by a factor of the number of frequencies. It is important to note that for systems without inversion symmetry, $\rho$ in Eq. \ref{capitalomega} and $V_{\rm xc}$ in Eqs. \ref{QP-offdiag} and \ref{QP-diag} are complex functions in reciprocal space (even though these are real functions when transformed to real space). For systems with inversion symmetry, $\Omega^2_{{\bf GG}'}{\left({\bf q}\;\!\right)}$ and $\tilde{\omega}^2_{{\bf GG}'}{\left({\bf q}\;\!\right)}$ are real, $\phi_{{\bf GG}'}{\left({\bf q}\;\!\right)}=0$ or $\pi$ and Eqs. \ref{gpp-sx} -- \ref{phi} reduce to a simpler form \cite{hybertsen86}.

In computing the sums in Eqs. \ref{gpp-sx} and \ref{gpp-ch} we drop terms in certain circumstances to save time and improve numerical precision. We neglect the terms for which $\left|\delta_{{\bf G}{\bf G}'}{-}\epsilon_{{\bf G}{\bf G}'}^{-1}({\bf q};0)\right|$, $\left| \lambda_{{\bf GG}'}{\left({\bf q}\;\!\right)} \right|$ or $\left|\cos\phi_{{\bf GG}'}{\left({\bf q}\;\!\right)}\right|$ are less than a given tolerance, since these terms have a vanishing contribution to the matrix elements of the self energy. This avoids ill-conditioned limits due to some of the intermediate quantities here being undefined. Another case is when for an occupied state $n''$,
$E\,{-}\,E_{n''{\bf k}{-}{\bf q}}{-}\,\tilde{\omega}_{{\bf GG}'}{\left({\bf q}\;\!\right)} \approx 0$,
in which case the GPP factors in $\Sigma_{\rm SX}$ and $\Sigma_{\rm CH}$ each diverge, although the sum
\begin{equation}
- \delta_{{\bf GG}'} + \frac{\Omega^2_{{\bf GG}'}{\left({\bf q}\;\!\right)}\left(1\,{-}\,i\tan\phi_{{\bf GG}'}{\left({\bf q}\;\!\right)}\right)}{2 \tilde{\omega}_{{\bf GG}'} \left(E\,{-}\,E_{n''{\bf k}{-}{\bf q}}{+}\,\tilde{\omega}_{{\bf GG}'}{\left({\bf q}\;\!\right)}\right)}
\end{equation}
remains finite. In this situation, we do not calculate these terms in $\Sigma_{\rm SX}$ and $\Sigma_{\rm CH}$ separately, but assign the sum of the contributions to $\Sigma_{\rm SX}$. When $n''$ is unoccupied there is only a $\Sigma_{\rm CH}$ contribution which diverges. Similarly, there are divergent contributions to $\Sigma_{\rm SX}$ when $E\,{-}\,E_{n''{\bf k}{-}{\bf q}}{+}\,\tilde{\omega}_{{\bf GG}'}{\left({\bf q}\;\!\right)} \approx 0$. In the full-frequency integrals in Eqs. \ref{ff-sx} and \ref{ff-ch}, we can see that the contributions around a pole of $\epsilon_{{\bf GG}'}^{-1}$ in this case vanish, so the correct analytic limit of these terms is zero \cite{hybertsen86}.

For static COHSEX calculations, the expressions used in the code are:
%\begin{widetext}
\begin{align}
\label{cohsex-sx}
\left<n{\bf k}\right|\Sigma_{\rm SX}{\left(0\right)}\left|n'{\bf k}\right>&=
-\sum_{n''}^{\rm occ}\sum_{{\bf qGG}'}
M^{*}_{n''n}({\bf k},-{\bf q},-{\bf G})M_{n''n'}({\bf k},-{\bf q},-{\bf G}') %\\
%\left<n{\bf k}\right|e^{i({\bf q}+{\bf G})\cdot{\bf r}}\left|n''{\bf k}{-}{\bf q}\right>
%\left<n''{\bf k}{-}{\bf q}\right|e^{-i({\bf q}+{\bf G}')\cdot{\bf r}'}\left|n'{\bf k}\right>\\
\\ \nonumber
&\times
\epsilon_{{\bf GG}'}^{-1}{\left({\bf q}\;\!;0\right)}
v{\left({\bf q}{+}{\bf G}'\right)}
\end{align}
and
\begin{align}
\label{cohsex-ch-sum}
\left<n{\bf k}\right|\Sigma_{\rm CH}&{\left(0\right)}\left|n'{\bf k}\right>=
\frac{1}{2}\sum_{n''}\sum_{{\bf qGG}'}
M^{*}_{n''n}({\bf k},-{\bf q},-{\bf G})M_{n''n'}({\bf k},-{\bf q},-{\bf G}') %\\
%\left<n{\bf k}\right|e^{i({\bf q}+{\bf G})\cdot{\bf r}}\left|n''{\bf k}{-}{\bf q}\right>
%\left<n''{\bf k}{-}{\bf q}\right|e^{-i({\bf q}+{\bf G}')\cdot{\bf r}'}\left|n'{\bf k}\right>\\
\\ \nonumber
&\times
\left[\epsilon_{{\bf GG}'}^{-1}{\left({\bf q}\;\!;0\right)}
\,{-}\,\delta_{{\bf GG}'}\right]
v{\left({\bf q}{+}{\bf G}'\right)} \\
%\end{align}
%or
%\begin{align}
\label{cohsex-ch-exact}
%\left<n{\bf k}\right|\Sigma_{\rm CH}{\left(0\right)}\left|n'{\bf k}\right>=
&=\frac{1}{2}\sum_{{\bf qGG}'}
M_{nn'}({\bf k},{\bf q}={\bf 0},{\bf G}'-{\bf G})
%\left<n{\bf k}\right|e^{i({\bf G}-{\bf G}')\cdot{\bf r}}\left|n'{\bf k}\right>
\left[\epsilon_{{\bf GG}'}^{-1}{\left({\bf q}\;\!;0\right)}
\,{-}\,\delta_{{\bf GG}'}\right]
v{\left({\bf q}{+}{\bf G}'\right)}
\end{align}
%\end{widetext}
where Eqs. \ref{cohsex-sx} and \ref{cohsex-ch-sum} can be derived formally from Eqs. \ref{gpp-sx} and \ref{gpp-ch} by setting ${\left(E\,{-}\,E_{n''{\bf k}{-}{\bf q}}\right)}$ to zero. Using the completeness relation for the sum over empty states, Eq. \ref{cohsex-ch-sum} can be written in a closed form given by Eq. \ref{cohsex-ch-exact}, which now does not involve the empty orbitals.

For Hartree-Fock calculations, we compute the matrix elements of bare exchange:
%\begin{widetext}
\begin{align}
\label{hf-x}
\left<n{\bf k}\right|\Sigma_{\rm X}\left|n'{\bf k}\right>&=
-\sum_{n''}^{\rm occ}\sum_{{\bf qGG}'}
M^{*}_{n''n}({\bf k},-{\bf q},-{\bf G})M_{n''n'}({\bf k},-{\bf q},-{\bf G}') %\\
%\left<n{\bf k}\right|e^{i({\bf q}+{\bf G})\cdot{\bf r}}\left|n''{\bf k}{-}{\bf q}\right>
%\left<n''{\bf k}{-}{\bf q}\right|e^{-i({\bf q}+{\bf G}')\cdot{\bf r}'}\left|n'{\bf k}\right>\\
%\nonumber
%&\times
\delta_{{\bf GG}'}
v{\left({\bf q}{+}{\bf G}'\right)}
\end{align}
%\end{widetext}

In principle, the inner and outer orbitals used in Eqs. \ref{ff-sx} -- \ref{hf-x} originate from the same mean-field solution. However, there is an option in the \texttt{sigma} executable to use a different mean-field solution for the inner and outer states. This is useful if one wishes to construct the $\Sigma$ operator within one mean field but expand the $\Sigma$ matrix using different orbitals, \textit{i.e.}, in order to evaluate matrix elements in a different basis than the mean-field wavefunctions when the quasiparticle wavefunctions are significantly different. This is also useful for verifying the accuracy of the linearization approximation as given by Eq. \ref{linearization} below.

Eq. \ref{QP-diag} depends on the evaluation energy parameter $E$. This parameter should be the quasiparticle energy $E^{\rm QP}_{n{\bf k}}$, determined self-consistently. In principle, what one may do is start by setting $E=E^{\rm MF}_{n{\bf k}}$ and find $E^{0}_{n{\bf k}}$ using Eq. \ref{QP-diag}. One can then set $E=E^{0}_{n{\bf k}}$ and solve Eq. \ref{QP-diag}, arriving at a new quasiparticle energy $E^{1}_{n{\bf k}}$. One can then repeat this process until convergence is reached. This process can be achieved using the different set of inner and outer states as described in the previous paragraph -- where the outer-state eigenvalues are updated after each step, and the eigenfunctions are left unchanged. In many cases, one can avoid this process by computing $\Sigma(E)$ on a grid of energies and interpolating or extrapolating to $E^{\rm QP}_{n{\bf k}}$. In particular, in many systems, $\Sigma(E)$ is a nearly linear function of $E$ so one may compute $\Sigma(E)$ for two grid points and evaluate the self-consistent $E^{\rm QP}_{n{\bf k}}$ using Newton's method \cite{hybertsen86}:
\begin{equation}
\label{linearization}
E^{\rm QP}_{n{\bf k}} = E^{0}_{n{\bf k}} + \frac{d\Sigma/dE}{1-d\Sigma/dE}(E^{0}_{n{\bf k}} - E^{\rm MF}_{n{\bf k}})
\end{equation}
The derivative that appears here is also related to the quasiparticle renormalization factor:
\begin{align}
Z = \frac{1}{1-d\Sigma/dE}
\end{align}

For full-frequency calculations, Eq. \ref{ff-sx} and Eq. \ref{ff-ch} are evaluated on a frequency grid, $E$, (not to be confused with the frequency grid over which the integrals are carried out) specified by the user. One then has access directly to $\mathrm{Re}\ \Sigma(\omega)$ and to $\mathrm{Im}\ \Sigma(\omega)$, printed in the file \texttt{spectrum.dat}, which can be used to construct the spectral function:
\begin{align}
\label{spectrum}
&A_{\bf k}(\omega)=\frac{1}{\pi} \cdot \\
\nonumber
&\sum_n \frac{ \left| \mathrm{Im}\ \Sigma_{n{\bf k}}(\omega) \right| }{(\omega - E^{\rm MF}_{n{\bf k}} - \mathrm{Re}\ \Sigma_{n{\bf k}}(\omega) + V_{\rm xc}^{n{\bf k}})^2 + \left| \mathrm{Im}\ \Sigma_{n{\bf k}}(\omega) \right|^2},
\end{align}
where we are using the mean-field exchange-correlation matrix element $V_{\rm xc}^{n{\bf k}} = \left< n {\bf k} \left| V_{\rm xc} \right| n {\bf k} \right>$.
This quantity can be used to compare directly with the quasiparticle spectrum from photo-emission experiments and various other measurements of the band-structure. 
 
The plane-wave matrix elements required in Eqs. \ref{ff-sx} -- \ref{hf-x} are similar to those of Eq. \ref{eqn_matrix_elem} required for the construction of the irreducible polarizability matrix. In the current case, however, we require additional matrix elements between valence-valence band pairs as well as conduction-conduction band pairs.  As was the case in the \texttt{epsilon} executable, the matrix elements are computed using FFTs utilizing the FFTW library \cite{fftw}. For each pair of outer states, $n$ and $n'$, we sum over all occupied and unoccupied inner states, $n''$, included in the calculation (typically states of energy up to the dielectric energy cutoff). Therefore, the computational cost of computing all the necessary matrix elements scales as $N^2 \log N$, where $N$ is the number of atoms (a factor of $N \log N$ comes from the FFTs). If one is interested in all the diagonal matrix elements, Eq. \ref{QP-diag}, in a given energy range (as opposed to just a fixed small number of states -- \textit{e.g.} VBM and CBM) then an additional factor of $N$ is included in the scaling which becomes $N^3 \log N$. If one requires both diagonal and off-diagonal elements within a given energy window (such as in a self-consistent GW scheme), then the scaling becomes $N^4 \log N$. 

Once the plane-wave matrix elements have been computed, the summations in the Coulomb-hole terms of Eqs. \ref{ff-sx} -- \ref{hf-x} for a particular $n$, $n'$ pair scale individually as $N^3$. Again, if all diagonal or off-diagonal matrix elements of $\Sigma$ in a given energy window must be computed an additional factor of $N$ or $N^2$ respectively is added to the scaling. 

\begin{figure}
\begin{center}
\includegraphics[width=6.324cm] {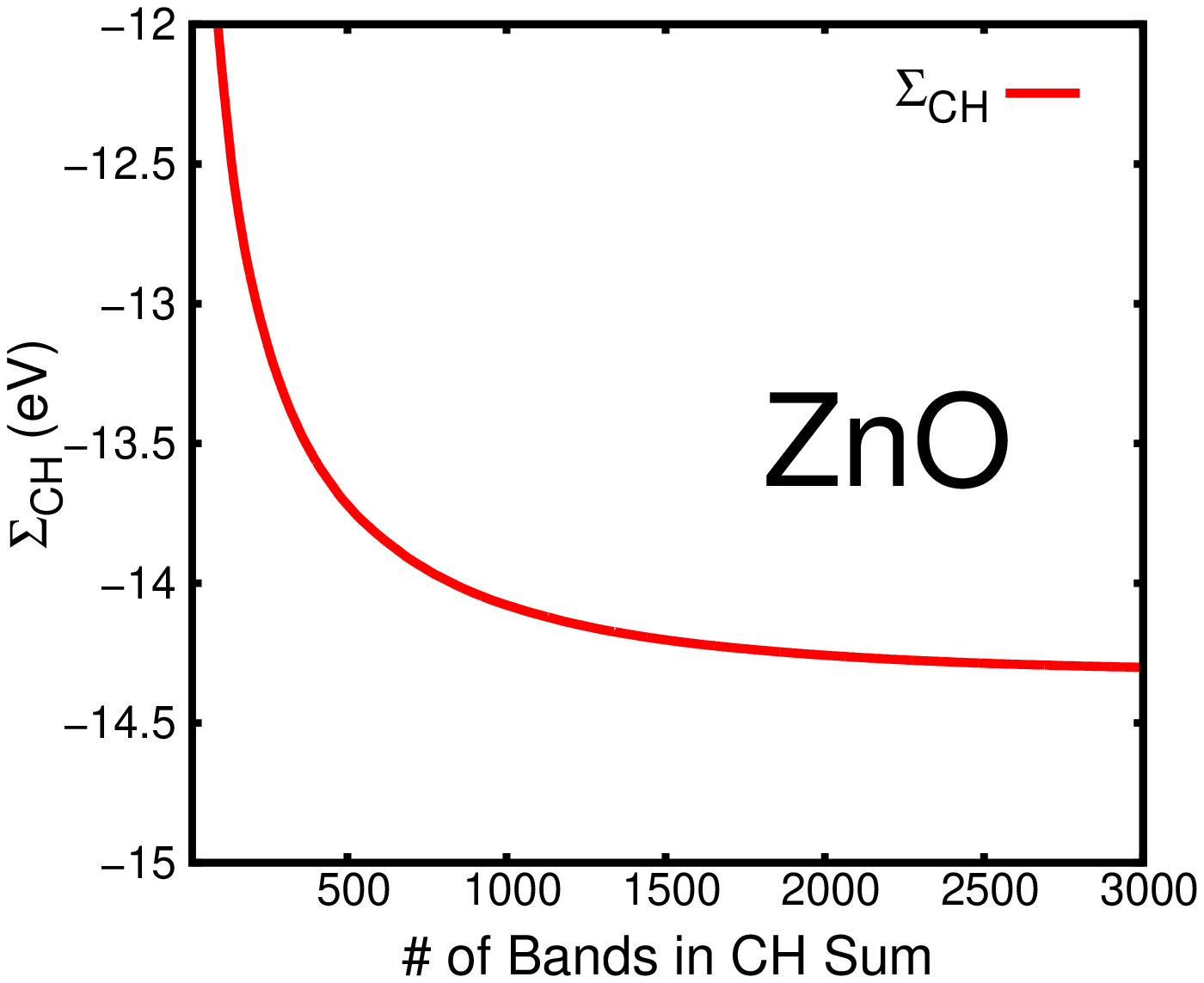}
\includegraphics[width=6.324cm] {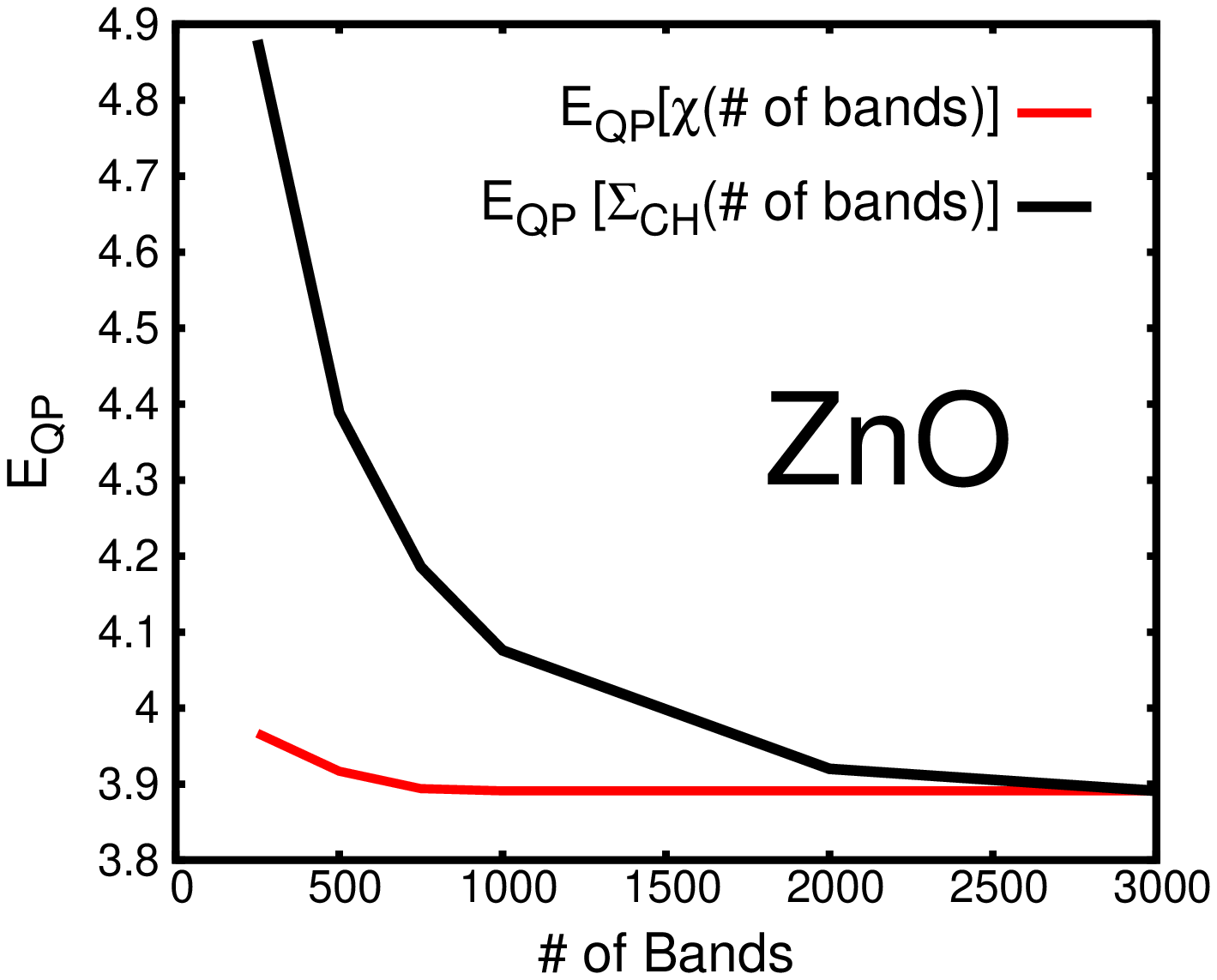}
\end{center}
\caption{ZnO Convergence of the VBM within the Hybertsen and Louie GPP model. (Left panel) Example output from file \texttt{ch\_convergence.dat} showing the Coulomb-hole sum value \textit{vs.} the number of bands included in the sum. (Right panel) The convergence of $E_{\rm QP}$ with respect to empty states in the polarizability sum, Eq. \ref{eqn_static_xi}, and with respect to empty states in the Coulomb-hole sum, Eq. \ref{gpp-ch}. The red curve shows the VBM $E_{\rm QP}$ in ZnO using a fixed 3,000 bands in the Coulomb-hole summation and varying the number of bands included in the polarizability summation.  The black curve shows the VBM $E_{\rm QP}$ in ZnO using a fixed 1,000 bands in the polarizability summation and varying the number of bands included in the Coulomb-hole summation. (For interpretation of the references to color in this figure legend, the reader is referred to the web version of this article.)}
\label{ZnO}
\end{figure}

It is important to point out that the Coulomb-hole summations in terms of Eqs. \ref{ff-sx} -- \ref{hf-x} converge exceptionally slowly with respect to the number of empty states included in the sums. The highest empty state included should have energy of at least the dielectric energy cutoff. Additionally, the convergence of the sums in \ref{ff-sx} -- \ref{hf-x} should be tested with respect to the dielectric energy cutoff.  As shown in Figure \ref{ZnO}, the convergence with respect to the dielectric energy cutoff, and the corresponding number of empty states, is very slow in many cases. This problem is similar to convergence issues with respect to empty states in the \texttt{epsilon} executable, as was discussed above. However, one finds that in many cases (particularly for bulk systems) the final $E^{\rm QP}$ converges much more slowly with respect to the number of empty states in the Coulomb-hole expression than in the polarizability expression \cite{shih10} -- see Fig. \ref{ZnO} for a comparison of these two rates in ZnO when using the Hybertsen and Louie GPP model. The partial sums of the Coulomb-hole matrix elements with respect to number of states included in the sum is written to the file \texttt{ch\_converge.dat}. Example output from \texttt{ch\_converge.dat} is plotted in Fig. \ref{ZnO}.

%=============================================================================
%=============================================================================

\subsection{Optical Properties: \texttt{BSE}}
\label{sec:bse}

The optical properties of materials are computed in the Bethe-Salpeter equation (BSE) executables. Here the eigenvalue equation represented by the BSE, Eq. \ref{BSE}, is constructed and diagonalized yielding  the excitation energies and wavefunctions of the correlated electron-hole excited states.  There are two main executables: \texttt{kernel} and \texttt{absorption}.  In the former, the electron-hole interaction kernel is constructed on a coarse ${\bf k}$-point grid, and in the latter the kernel is (optionally) interpolated to a fine ${\bf k}$-point grid and diagonalized.

The \texttt{kernel} executable constructs the second term of the left-hand side of Eq. \ref{BSE} which is referred to as the electron-hole kernel. The kernel, $K$, as implemented in the package, is limited to the static approximation, and contains two terms, a screened direct interaction and a bare exchange interaction, $K^{\rm eh}=K^{\rm d}+K^{\rm x}$, defined in the following way \cite{rohlfing00}:
\begin{align}
\label{kdirect}
\langle vc{\bf k} | K^{\rm d} &| v' c' {\bf k}' \rangle = \\ 
\nonumber
-&\int d{\bf r} d{\bf r}' \psi^{*}_{c {\bf k}}({\bf r})\psi_{c' {\bf k}'}({\bf r})W({\bf r},{\bf r}')\psi^{*}_{v' {\bf k}'}({\bf r}')\psi_{v {\bf k}}({\bf r}')
\end{align}
and
\begin{align}
\label{kexchange}
\langle vc{\bf k} | K^{\rm x} &| v' c' {\bf k}' \rangle = \\ 
\nonumber
&\int d{\bf r} d{\bf r}' \psi^{*}_{c {\bf k}}({\bf r})\psi_{v {\bf k}}({\bf r})v({\bf r},{\bf r}')\psi^{*}_{v' {\bf k}'}({\bf r}')\psi_{c' {\bf k}'}({\bf r}').
\end{align}
These matrices are constructed on a coarse grid of $\textbf{k}$-points, in most cases the same grid used within the GW calculation because one must have previously constructed the dielectric matrix $\epsilon^{-1}({\bf q})$ for ${\bf q}={\bf k}-{\bf k}'$. We calculate these matrices in $\textbf{G}$-space using the prescription of Rohlfing and Louie \cite{rohlfing00}:
\begin{align}
\label{direct_term}
\langle vc{\bf k} | K^{\rm d} &| v' c' {\bf k}' \rangle = \\ 
\nonumber
\sum_{{\bf GG}'}&M_{c'c}^{*}({\bf k},{\bf q},{\bf G})W_{{\bf G}{\bf G}'}({\bf q};0)M_{v'v}({\bf k},{\bf q},{\bf G}')
\end{align}
and
\begin{align}
\label{exchange_term}
\langle vc{\bf k} | K^{\rm x} &| v' c' {\bf k}' \rangle = \\ 
\nonumber
\sum_{{\bf G} \ne 0}&M_{vc}^{*}({\bf k},{\bf q},{\bf G})v({\bf q}+{\bf G})M_{v'c'}({\bf k},{\bf q},{\bf G})
\end{align}
where $M$ is defined in Eq. \ref{eqn_matrix_elem} and calculated using FFTs as described above in Sec. \ref{sec:xi0}. 

For each ${\bf k}$ and ${\bf k}'$, we must therefore calculate all the matrix elements $M_{vv'}$, $M_{cc'}$, and $M_{vc}$.  The number of valence and conduction bands required to calculate the absorption spectrum within a given energy window each scales linearly with the number of atoms $N$. So, formally we again have $N^3 \log N$ scaling with the use of FFTs.  The summations involved in Eq. \ref{direct_term} and Eq. \ref{exchange_term}, however, formally scale as $N^5$ since there are $N^4$ terms to compute and each involves a sum over ${\bf GG}'$ that may be done sequentially, first ${\bf G}'$ and then ${\bf G}$. In practice, though, except for the largest systems considered, the summations require less time than the matrix elements, and $N_v$ and $N_c$ remain small compared to the values required in the GW step, for example where states with energy up to the dielectric cutoff were required. Usually the energy window used in solving the BSE is approximately 10 eV, giving a spectrum converged beyond the visible region. As we discuss below, within the \texttt{BerkeleyGW} package, the parallel wall-time scales as $N$ for this step. However, the $N^5$ scaling will present a considerable challenge when applying the code to systems of size greater than 100s of atoms.

As was the case for GW code, the ${\bf q}\rightarrow0$ limit must be handled carefully and differently depending on the type of screening in the system. For the exchange kernel, we zero out all ${\bf G=G'}=0$ contributions to the kernel matrix elements, as discussed in Ref. \cite{hanke78} which gives directly $\mathrm{Im}\ \epsilon_{\rm M}$ where $\epsilon_{\rm M}$ is the macroscopic dielectric constant. For the direct term, however, we must handle the ${\bf G}=0$ case specially. For these purposes, the ${\bf G}=0$ and ${\bf G}'=0$ terms are removed from Eq. \ref{direct_term} and treated separately. For each $({\bf k}cv,{\bf k}'c'v')$ we save three terms: the body term, which contains the result of the sum in Eq. \ref{direct_term} with the ${\bf G}=0$ or ${\bf G}'=0$ terms removed; the wing term, which contains all the sum of all the remaining terms in the sum with the exception of the single term where ${\bf G=G}'=0$; the head term, which contains the remaining term from the sum where ${\bf G=G}'=0$. For metallic systems, as we discussed above, $\epsilon^{-1}({\bf q},{\bf G}={\bf G}'=0) \propto 1/v({\bf q})$ so that $W({\bf q},{\bf G}={\bf G}'=0) \propto C$, thus the head term in the kernel remains well behaved. For semiconductors however, $W({\bf q},{\bf G}={\bf G}'=0) \propto 1/q^2$ so that the head term in the kernel actually diverges as $1/q^2$ when ${\bf q}\rightarrow0$. Similarly, the wing term diverges as $1/q$ when ${\bf q}\rightarrow0$ for semiconductors, while it again remains well behaved for metals. These limits are summarized in Table \ref{w_table}.

Because exciton binding energies and absorption spectra depend sensitively on quantities like the joint density of states, it is essential in periodic systems to sample the ${\bf k}$-points on a very fine grid. Directly calculating the kernel on this fine grid in the \texttt{kernel} executable would be prohibitively expensive, so instead we interpolate the kernel in the \texttt{absorption} executable before diagonalization. For semiconductors, the head and wing kernel terms are not smooth functions of ${\bf k}$ and ${\bf k'}$ (as we have shown above, they diverge for ${\bf q}={\bf k}-{\bf k}'\rightarrow0$). Therefore, the quantities that we interpolate are $q^2 \cdot K^{\rm d}_{\text{head}}$, ${\bf q} \cdot K^{\rm d}_{\text{wing}}$ and the body term directly as they are now smooth quantities \cite{rohlfing00}. For metals, we interpolate directly the kernel without any caveats because all the contributing terms are smooth functions of ${\bf k}$ and ${\bf k'}$. As in GW, we treat metals with zero-temperature occupations.

The \texttt{absorption} executable requires both coarse- and fine-grid wavefunctions as input. The interpolation is done through a simple expansion of the fine-grid wavefunction in terms of nearest coarse-grid wavefunction:
\begin{equation}
\label{kpexpansion}
u_{n{\bf k}_{\text{fi}}}=\sum_{n'} C_{n,n'}^{{\bf k}_{\text{co}}} u_{n'{\bf k}_{\text{co}}}
\end{equation}
where ${\bf k}_{\text{co}}$ is the closest coarse-grid point to the fine-grid point, ${\bf k}_{\text{fi}}$, and the coefficients $C_{n,n'}^{{\bf k}_{\text{fi}}}$ are defined as the overlaps between the coarse-grid and fine-grid wavefunctions:
\begin{equation}
\label{coefficients}
C_{n,n'}^{{\bf k}_{\text{co}}} = \int d{\bf r} \,u_{n{\bf k}_{\text{fi}}}({\bf r}) u^{*}_{n'{\bf k}_{\text{co}}}({\bf r}).
\end{equation}
The coefficients $C_{n,n'}^{{\bf k}_{\text{co}}}$ are normalized so that $\sum_{n'} |C_{n,n'}^{{\bf k}_{\text{co}}}|^2=1$. It should be noted that for a given set of fine bands one can improve the interpolation systematically by including more valence and conduction bands in the coarse grid due to the completeness of the Hilbert space at each ${\bf k}$. It should also be noted that we do restrict $n$ and $n'$ to be either both valence or both conduction bands -- this is acceptable due to the different character of the conduction and valence bands in most systems.

Using these coefficients we interpolate the kernel with the following formula:
\begin{align}
\label{interpolation}
\langle &vc{\bf k}_{\text{fi}} | K | v' c' {\bf k}'_{\text{fi}} \rangle = \\ 
\nonumber
&\sum_{n_1,n_2,n_3,n_4} C_{c,n_1}^{{\bf k}_{\text{co}}} C_{v,n_2}^{*{\bf k}_{\text{co}}} C_{c',n_3}^{*{\bf k}'_{\text{co}}} C_{v',n_4}^{{\bf k}'_{\text{co}}} \langle n_2 n_1 {\bf k}_{\text{co}} | K | n_4 n_3 {\bf k}'_{\text{co}} \rangle
\end{align}
where $K$ is one of the head, wing, body or exchange kernel terms. As in the case of \texttt{epsilon}, this summation can be performed compactly as a set of matrix-matrix multiplications. We utilize the Level 3 BLAS calls \texttt{DGEMM} and \texttt{ZGEMM} to optimize the performance.

\begin{figure}
\begin{center}
\includegraphics[width=6.324cm] {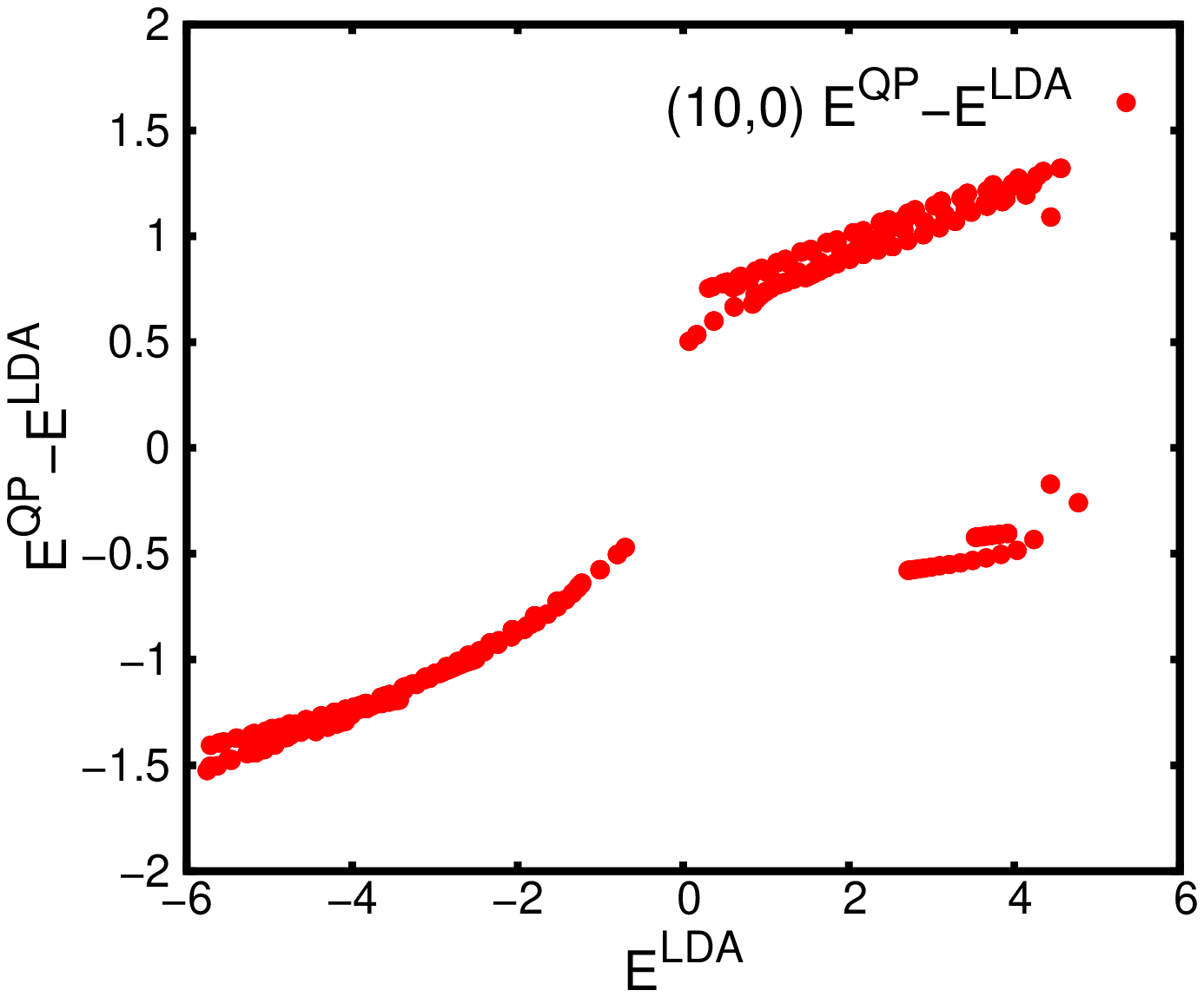} \\
\includegraphics[width=6.324cm] {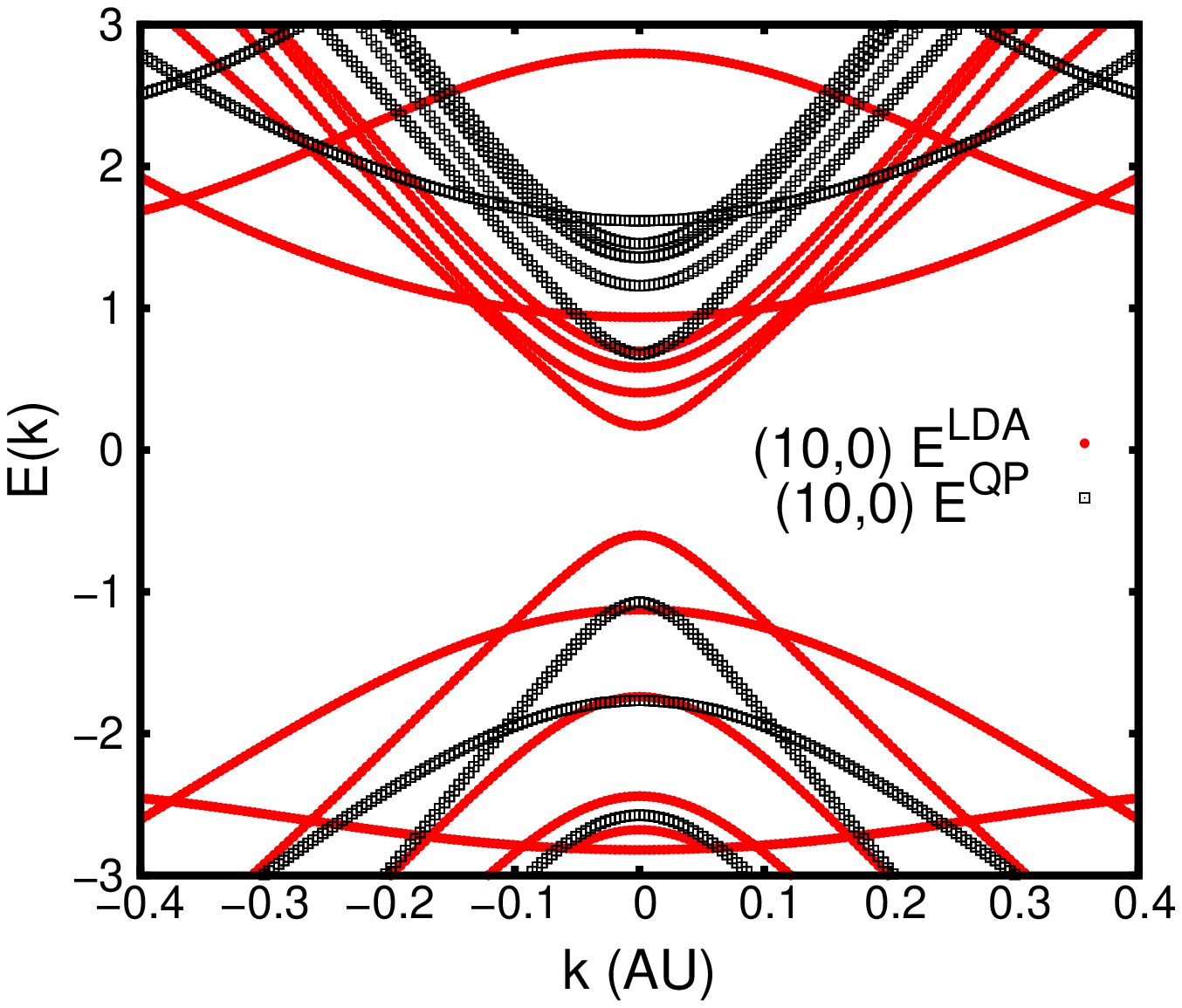}
\end{center}
\caption{Top: GW quasiparticle self-energy corrections, $E^{\rm QP} - E^{\rm LDA}$ \textit{vs.} the LDA energy for (10,0) SWCNT. Both a rigid opening of the band gap and a non-linear energy scaling are present. Bottom: The fine-grid quasiparticle band-structure using the interpolated self-energy corrections (black open) and the LDA uninterpolated band-structure (red). 256 points are used to sample the Brillouin zone in the Coulomb-hole summation. (For interpretation of the references to color in this figure legend, the reader is referred to the web version of this article.)}
\label{10-0}
\end{figure}

One can improve on the interpolation systematically by using the closest four coarse-grid points to each fine point and using a linear interpolation layer in addition to the wavefunction-based interpolation described above.  This is done by default for the interpolation of the first term of Eq. \ref{BSE} for the quasiparticle self-energy corrections $E^{\rm QP}-E^{\rm MF}$: 
\begin{align}
\label{kpeig}
E^{\rm QP}_n&({\bf k}_{\text{fi}}) = \\ 
\nonumber
&E^{\rm MF}_n({\bf k}_{\text{fi}}) + \left< \sum_{n'} \left| C_{n,n'}^{{\bf k}_{\text{co}}} \right|^2 \left(E^{\rm QP}_{n'}({\bf k}_{\text{co}}) - E^{\rm MF}_{n'}({\bf k}_{\text{co}}) \right) \right>_{{\bf k}_{\text{co}}}
\end{align}
where the brackets indicate linear interpolation using the tetrahedron method. In this case, the wavefunction-based interpolation layer guarantees that the band crossings are properly handled, and the linear interpolation layer ensures that we correctly capture the energy dependence of the self-energy corrections. In this way, we can construct $E^{\rm QP}$ on the fine grid, or any arbitrary point, given $E^{\rm MF}$ on the fine grid and $E^{\rm QP}$ and $E^{\rm MF}$ on the coarse grid (Fig. \ref{10-0}).

As an alternative to calculating the quasiparticle corrections on the coarse grid and interpolating them to the fine grid, the user may choose a less refined method of specifying the corrections using a three-parameter model involving a scissor-shift parameter $\Delta E$ to open the energy gap at the Fermi energy, a zero energy $E_0$ (typically the band edge), and an energy-scaling parameter $C$ changing the bandwidth (the parameters are specified separately for valence and conduction bands):
\begin{align}
E^{\rm QP} = E^{\rm MF} + \Delta E + C \left( E^{\rm MF} - E_0 \right).
\end{align}

Having constructed the kernel on the fine grid, we now consider the diagonalization of the kernel. The kernel matrix is of dimension $N_c \cdot N_v \cdot N_k$ where $N_k$ is the number of ${\bf k}$-points on the fine grid.  Formally, the kernel dimension scales as $N$ for periodic systems with small unit cells and $N^2$ for large systems, where $N$ is the number of atoms. For bulk systems with small unit cells, $N_k \propto 1/N$, but the reduction with increased cell size saturates quickly for large systems and large molecules where $N_k=O(1)$ (to compute a smooth continuum absorption onset, it is necessary to include some level of $\textbf{k}$-point sampling even for isolated systems). The matrix can be diagonalized exactly within LAPACK (\texttt{zheevx}) or ScaLAPACK (\texttt{pzheevx}). The diagonalization therefore scales as $N^3$ for periodic systems with small unit cells and $N^6$ for large systems.

The result of the diagonalization is the set of exciton eigenvalues $\Omega^S$ and eigenfunctions $A^{S}_{cv{\bf k}}$ which can be used to construct the joint density of states (JDOS) or the absorption spectrum (or ${\rm Im}\ \epsilon_2 \left( \omega \right)$) using Eq. \ref{eps2}. There are a number of post-processing tools in the package, such as \texttt{PlotXct}, which plots the exciton wavefunction in real space according to Eq. \ref{plotxct} to analyze the exciton states. The \texttt{absorption} executable can produce both the singlet and triplet eigenvalues and eigenvectors.  In the latter case, the exchange term is set to zero when diagonalizing the kernel \cite{rohlfing00}. (Note that for triplets the oscillator strengths calculated in the code are evaluated without considering spin overlap. In some cases this ``oscillator strength of the corresponding singlet'' may be useful. The true physical oscillator strength of course is zero for triplets.)  Additionally, one may compute the eigenvalues and eigenvectors with only the exchange interaction; the resulting spectrum should be the same as the one obtained within RPA with local-field effects included \cite{hanke78}.

The $N^6$ scaling for large systems in the diagonalization is, in practice, much more limiting than the $N^5$ step in the construction of the kernel. This is because the latter step can be parallelized very efficiently, while diagonalization, even with the use of ScaLAPACK, typically saturates at $O(1000)$ CPUs. Often one is only interested in the absorption spectrum or the JDOS, and not all of the correlated exciton eigenfunctions and eigenvalues. For such systems, we use the Haydock recursion method \cite{haydock80,benedict99}. This is an iterative method based on spectral decomposition and requires only matrix-vector products, which can be parallelized efficiently. This method gives the absorption spectrum directly, the equivalent of Eq. \ref{eps2}. In principle, one can get eigenvalues and eigenvectors for a small energy range of interest using iterative Lanczos algorithms \cite{benedict99}.  

As mentioned above, the electron-hole kernel should be constructed with a sufficient number of valence and conduction bands to cover the energy window of interest -- typically all bands within the desired energy window from the Fermi energy should be included so that the energy window of the bands included in the calculation is at least twice that of the desired absorption energy window. The \texttt{absorption} executable computes the percent deviation from the $f$-sum rule \cite{hybertsen86}:
\begin{equation}
\int_{0}^{\infty} \epsilon_2(\omega) \omega d\omega = - \frac{\pi \omega_{\rm p}^2}{2}.
\end{equation}
One should converge this quantity with both the number of valence and conduction bands included. The absorption spectrum (or $\epsilon_2$) in the energy window of interest converges much more quickly than $\epsilon_1$ if high-energy transitions outside of the window of interest contribute greatly to the sum rule, since $\epsilon_1(\omega)$ is related to an integration over all frequencies of $\epsilon_2(\omega)$ via the Kramers-Kronig relation.  

Finally, the transition matrix elements $\left<0|{\bf v}|S\right>$ in Eq. \ref{eps2} are printed in the file \texttt{eigenvalues.dat}. These are related to the oscillator strengths $f_S$ by
\begin{eqnarray}
f_S = \frac{2 \left|{\bf e} \cdot \left<0|{\bf v}|S\right> \right|^2}{\Omega^S}
\end{eqnarray}
We compute the velocity matrix element via the commutator of the many-body Hamiltonian, as follows \cite{tiagothesis}:
\begin{eqnarray}
\left< 0 \left| {\bf v} \right| S \right> = \left< 0 \left|  i \left[ H, {\bf r} \right]  \right| S \right> = i \left( E_0 - E_S \right) \left< 0 \left|  {\bf r} \right| S \right> \nonumber \\
= -i \Omega^S \sum_{vc\bf k} A^S_{vc\bf k} \left< v {\bf k} \left| {\bf r} \right| c {\bf k} \right>
\end{eqnarray}
In a periodic system, we cannot calculate matrix elements of the position operator, but we can use a ${\bf q} \rightarrow 0$ limit \cite{rohlfing00}:
\begin{eqnarray}
 \left< v {\bf k} \left| {\bf r} \right| c {\bf k} \right> = \lim_{{\bf q} \rightarrow 0} \frac{\left< v {\bf k + q} \left| e^{i {\bf q} \cdot {\bf r}} - 1 \right| c {\bf k} \right>}{iq} \nonumber \\
= -i \lim_{{\bf q} \rightarrow 0} \frac{\left< v {\bf k + q} \left| e^{i {\bf q} \cdot {\bf r}}\right| c {\bf k} \right>}{q}
\label{velocityop}
\end{eqnarray}
In practice, we evaluate the limit using finite differences for a small value of ${\bf q}$, similarly to how the limit is treated in the \texttt{Epsilon} code. Thus the valence bands on a shifted fine ${\bf k}$-grid are required. (Note we are assuming an interband transition.)

As an alternative to the finite-difference approach, one may approximate the velocity operator by the momentum operator $-i \nabla$, avoiding the calculation of the valence bands on the shifted fine grid. We reverse the derivation partly:
\begin{eqnarray}
\left< v {\bf k} \left| {\bf r} \right| c {\bf k} \right> = \frac{\left< v {\bf k} \left| \left[ H^{\rm MF}, {\bf r} \right] \right| c {\bf k} \right>}{E^{\rm MF}_{v {\bf k}} - E^{\rm MF}_{c {\bf k}}} =
-i \frac{\left< v {\bf k} \left| {\bf v}^{\rm MF} \right| c {\bf k} \right>}{E^{\rm MF}_{v {\bf k}} - E^{\rm MF}_{c {\bf k}}} \approx - \frac{\left< v {\bf k} \left| \nabla \right| c {\bf k} \right>}{E^{\rm MF}_{v {\bf k}} - E^{\rm MF}_{c {\bf k}}}
\end{eqnarray}
This does not require an additional grid, but yields inexact oscillator strengths due to neglect of commutators between ${\bf r}$ and the non-local part of the Hamiltonian \cite{rohlfing00,sohrab01}. We could have approximated the quasiparticle or excitonic velocity operator by the momentum operator, but this would be less accurate, since those Hamiltonians have additional sources of non-locality beyond those of the mean-field Hamiltonian.
The momentum operator uses transitions $v {\bf k} \rightarrow c {\bf k}$ in the Bethe-Salpeter equation (Eq. \ref{BSE}), whereas with the velocity operator we actually use the transitions $v {\bf k} + {\bf q} \rightarrow c {\bf k}$. This ensures consistency of wavefunctions between excitons and transition matrix elements, and also is needed to describe transverse and longitudinal excitons in materials with an indirect gap \cite{rohlfing00}.

%============================================================================================
%============================================================================================

\section{Parallelization and Performance}

\subsection{\texttt{epsilon}}
\label{sec:xi0_parallel}

The parallelization of \texttt{epsilon} is characterized by two distinct schemes for the two main sections of the code: (1) the computation of matrix elements (Eq. \ref{eqn_matrix_elem}), and (2) the matrix multiplication (Eq. \ref{eqn_static_xi_compact}) and inversion.

For the computation of the matrix elements in Eq. \ref{eqn_matrix_elem}, the code is parallelized with nearly linear scaling up to $N_v \cdot N_c$ processors, where $N_v$ and $N_c$ are the number of valence and conduction bands respectively used in the sum of Eq. \ref{eqn_static_xi}. Each processor owns an approximately equal fraction of the total number of $(v,c)$ pairs for all {\bf k}, and performs serial FFTs to compute the matrix elements, Eq. \ref{eqn_matrix_elem}, for all ${\bf G}$ and ${\bf k}$ associated with the pair.  Note that for large systems, $N_v$ is on the order of 100s and $N_c$ is on the order of 1000s or more, so that this section of the code scales well up to 100,000 CPUs. 

All wavefunctions are stored in memory unless the optional \texttt{comm\_disk} flag is given. Each processor holds in memory the wavefunctions for all the pairs it owns. If \texttt{comm\_disk} is specified (as opposed to the default \texttt{comm\_mpi} option), the distribution of pairs is the same, but each processor saves the conduction wavefunctions it needs on disk and reads the wavefunctions back into memory one pair at time for the purposes of computation. Using \texttt{comm\_disk} can therefore reduce the amount of memory required for the computation, but comes with a substantial performance reduction.

\begin{figure}
\begin{center}
\includegraphics[width=6.324cm] {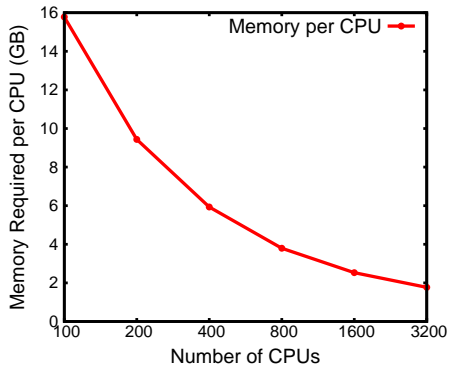}
\end{center}
\caption{The memory required per CPU \textit{vs.} the number of CPUs used for a \texttt{epsilon} calculation on the (20,20) nanotube. See text for parameters used.}
\label{xi0Memory}
\end{figure}

The processors are distributed into valence and conduction band pools in order to minimize the memory required using a complete search algorithm. For example, if one calculates a material with few valence and many conduction bands, all the processors will be in one or two valence pools (holding all or half the valence bands in memory each) but spread over a large number of conduction pools because the relative cost of holding all the valence bands in memory is much smaller than holding all the conduction bands in memory. In such a scheme, the amount of memory required per processor drops linearly with small numbers of processors and decreases as $1/\sqrt{N_{\rm proc}}$ for large numbers of processors (Fig. \ref{xi0Memory}).

In the second section of the \texttt{epsilon} code, we switch from a parallelization over bands to a parallelization scheme over ${\bf GG}'$ for the polarizability (Eq. \ref{eqn_static_xi}) and dielectric matrices (Eq. \ref{epsilon}). We use the ScaLAPACK block-cyclic layout \cite{scalapack} in anticipation of utilizing the ScaLAPACK libraries for the inversion of the dielectric matrix. The transition between the band distribution of the matrix elements in Eq. \ref{eqn_matrix_elem} and the block-cyclic layout of the polarizability matrix is achieved naturally in the process of doing the parallel matrix-matrix multiplication involved in Eq. \ref{eqn_static_xi_compact}. There is, however, a significant amount of communication involved at this step. To minimize this communication we have two options for the parallel multiplication.

In the first scheme, corresponding to the use of the \texttt{gcomm\_matrix} flag in \texttt{epsilon.inp}, we loop over processors (for simplicity, we label the loop index $i$) who own a piece of the polarizability matrix $\chi({\bf G},{\bf G}')$. Each processor does the fraction of the matrix multiplication in Eq. \ref{eqn_static_xi_compact} relating to the $(n,n')$ pairs it owns and for submatrix $\chi({\bf G}_i,{\bf G}_i')$ that the $i$th processor stores. The processors then MPI-reduce their contribution to the $i$th processor. Thus the total communication in this scheme is an eventual reduction of the entire $\chi({\bf G},{\bf G}')$ to the processors that store it. It is important to note that we must do this one processor at time (or at most in chunks of processors -- chosen often as the number of CPUs per node) because no single processor can hold in memory the entire $\chi({\bf G},{\bf G}')$ matrix for large systems.

In the second scheme, corresponding to the use of the \texttt{gcomm\_elements} flag in \texttt{epsilon.inp}, we again loop over processors, but this time, we have the $i$th processor MPI-broadcast to all processors that hold a piece of the polarizability matrix the set of matrix elements for all the $(v,c)$ pairs it owns.  Each processor then uses these matrix elements to compute the contribution of the matrix-matrix product, Eq. \ref{eqn_static_xi_compact}, for the submatrix of $\chi({\bf G},{\bf G}')$ it stores.  In this scheme, all the matrix elements (Eq. \ref{eqn_matrix_elem}) are eventually broadcast.  

Whether the use of \texttt{gcomm\_elements} or \texttt{gcomm\_matrix} is optimal depends on whether it is faster to reduce the $\chi({\bf G},{\bf G}';\omega)$ or broadcast all the matrix elements, $M_{nn'}({\bf k},{\bf q},\{{\bf G}\})$.  In particular if $N_{\bf G} \cdot N_{\rm freq} < N_v \cdot N_c \cdot N_k$, where $N_{\rm freq}$ is the number of frequencies in a full frequency calculation, then it is cheaper to use \texttt{gcomm\_matrix}. If no flag is specified, the \texttt{epsilon} code will make this choice for the user based on the above criteria.

\begin{figure}
\begin{center}
\includegraphics[width=6.324cm] {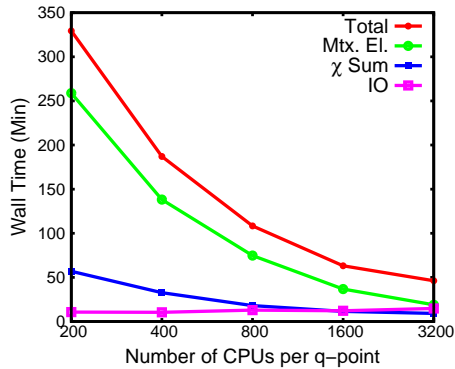}
\end{center}
\caption{The wall-time required \textit{vs.} the number of CPUs per ${\bf q}$-point used for a \texttt{epsilon} calculation on the (20,20) single-walled carbon nanotube. There is near linear scaling up to 1,600 CPUs. Since there is an additional layer of trivial parallelization over the 32 ${\bf q}$-points required, the \texttt{epsilon} calculation scales to over 50,000 CPUs. See text for parameters used.}
\label{xi0Time}
\end{figure}

Because we use the block-cyclic layout, the memory required to store the $\chi({\bf G},{\bf G}')$ decreases linearly with the number of CPUs.  However, the cost of the inversion utilizing ScaLAPACK can saturate at 100s of CPUs and the cost of the summation can saturate with a few thousand CPUs, see Figure \ref{xi0Time}. In general, the number of CPUs used for the block-cyclic distribution of $\chi$ can be tuned.

Beyond the more sophisticated level of parallelization described above, there is a more trivial level of parallelization available to small systems requiring large numbers of $\textbf{k}$-points: Eq. \ref{eqn_static_xi} is completely separable as a function of ${\bf q}$. One may run a separate \texttt{epsilon} calculation for each ${\bf q}$ required and merge the dielectric matrices -- in such a way, a user can obtain perfectly linear artificial scaling with CPUs to $N_{k}$ times the number CPUs mentioned above.

The scaling of memory and computation time with respect to the number of CPUs used per {\bf q}-point in \texttt{epsilon} for the example (20,20) SWCNT calculation is shown in Fig. \ref{xi0Memory} and Fig \ref{xi0Time}. We find nearly linear scaling up to 3200 CPUs per {\bf q}-point. Since there are 32 {\bf q}-points in this calculation that are trivially parallelized, we find nearly linear scaling of the \texttt{epsilon} computation up to $\sim$100,000 CPUs.

%%%%%%%%%%%%%%%%%%%%%%%%%%%%%%%%%%%%%%%%%%%%%%%%%%%%%%%%%%%%%%%%%%%%%%%%%%%%%%%%%%%%%%%%%%
%%%%%%%%%%%%%%%%%%%%%%%%%%%%%%%%%%%%%%%%%%%%%%%%%%%%%%%%%%%%%%%%%%%%%%%%%%%%%%%%%%%%%%%%%%

\subsection{\texttt{sigma}}
\label{sec:sigma_parallel}

Within a \texttt{sigma} calculation, one computes a requested number of diagonal, Eq. \ref{QP-diag}, or off-diagonal, Eq. \ref{QP-offdiag}, $\Sigma$ matrix elements. For each matrix element there are two computationally intensive steps.  The first is to calculate all the plane-wave matrix elements $M_{nn''}$ and $M_{n'n''}$, Eq. \ref{eqn_matrix_elem}, for the outer states of interest, $n$ and $n'$, and for all occupied and empty states, $n''$.  Secondly, we compute the sum over states, $n''$, as well as ${\bf G}$, ${\bf G}'$ and ${\bf q}$ in the expressions in Eqs. \ref{ff-sx} -- \ref{hf-x}.  

The \texttt{sigma} execution is parallelized over both outer bands, $n$ and $n'$, and inner bands, $n''$. As in the case of \texttt{epsilon}, this is done by defining pools and distributing the $n$, $n'$ pairs evenly among the pools. We then distribute the $n''$ bands evenly within the pools. As was the case for \texttt{epsilon}, we define the number of pools using a complete search algorithm to minimize the amount of memory per CPU required to store the inner and outer wavefunctions.

As described above, the CPU time required for the computation of all plane-wave matrix elements, $M_{nn''}$ and $M_{n'n''}$, scales as $N^2 \log N$, where $N$ is the number of atoms, for each $\Sigma$ matrix element of interest. As described above, the outer-state pairs are parallelized over pools and the inner states are parallelized over the CPUs within each pool. The wall-time for the computation of all the plane-wave matrix elements required for every $\Sigma$ matrix element scales as $N \log N$ with unlimited CPU resources. As was the case in the \texttt{epsilon} executable, each CPU computes the plane-wave matrix elements between all the $(n,n'')$ and $(n',n'')$ pairs it owns for all ${\bf G}$ through serial FFTs using FFTW \cite{fftw}.

The summations required in Eqs. \ref{ff-sx} -- \ref{hf-x} are parallelized by again distributing the outer-state pairs over the pools and then distributing the inner states over the CPUs within each pool. The wall-time for the summations, therefore, scales as $N^2$ (for the sums over ${\bf G}$ and ${\bf G}'$) regardless of the number of diagonal or off-diagonal elements requested, given unlimited CPU resources.

As was the case for \texttt{epsilon}, the wavefunctions are distributed in memory, with each CPU owning only the $n$, $n'$ and $n''$ wavefunctions that it needs for the computations described above.  The dielectric matrix, $\epsilon^{-1}_{{\bf G},{\bf G}'}({\bf q};E)$ for each ${\bf q}$ and $E$, is distributed globally over the matrix rows, ${\bf G}$.

\begin{figure}
\begin{center}
\includegraphics[width=6.324cm] {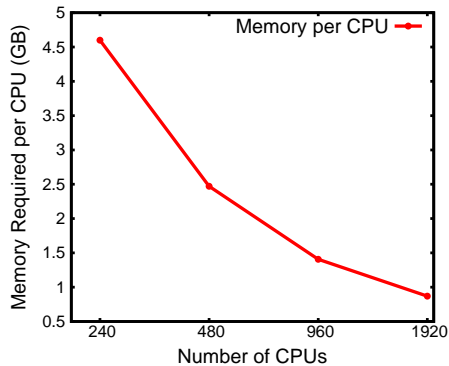}
\end{center}
\caption{The memory required per CPU \textit{vs.} the number of CPUs used for a \texttt{sigma} calculation on the (20,20) nanotube. See text for parameters used.}
\label{sigMemory}
\end{figure}

\begin{figure}
\begin{center}
\includegraphics[width=6.324cm] {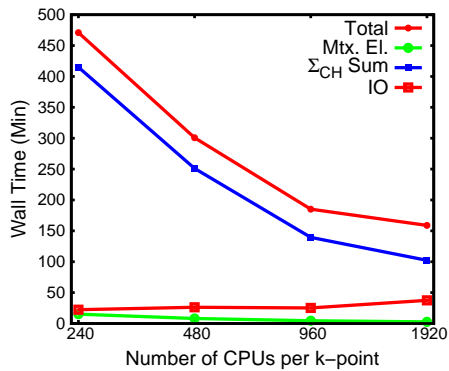}
\end{center}
\caption{The wall-time required \textit{vs.} the number of CPUs per ${\bf k}$-point used for a \texttt{sigma} calculation on the (20,20) single-walled carbon nanotube. There is near linear scaling up to 1,920 CPUs. Since there is an additional layer of trivial parallelization over the 16 ${\bf k}$-points required, the \texttt{sigma} calculation scales to over 30,000 CPUs. See text for parameters used.}
\label{sigTime}
\end{figure}

The scaling of memory and computation time with respect to the number of CPUs used per {\bf k}-point in \texttt{sigma} for the example (20,20) SWCNT calculation is shown in Fig. \ref{sigMemory} and Fig \ref{sigTime}. We find nearly linear scaling up to 1600 CPUs per {\bf k}-point. Since there are 16 irreducible {\bf k}-points in this calculation that are trivially parallelized, we find nearly linear scaling of the \texttt{sigma} computation up to ~25,000 CPUs.

\subsection{\texttt{BSE}}
\label{sec:bse_parallel}

As mentioned in the previous sections, in the \texttt{kernel} executable, for each ${\bf k}$ and ${\bf k}'$, we must calculate all the matrix elements $M_{vv'}$, $M_{cc'}$, and $M_{vc}$ and then perform the summations involved Eq. \ref{direct_term} and Eq. \ref{exchange_term} for each $(vc{\bf k},v'c'{\bf k}')$ pair. \texttt{BerkeleyGW} automatically  parallelizes this in different schemes depending on the system and number of CPUs provided.  

If the number of CPUs is less than $N_k^2$, the square of the number of coarse $\textbf{k}$-points, we distribute the $({\bf k},{\bf k}')$ pairs evenly over the CPUs and each CPU calculates all the matrix elements,  $M_{vv'}$, $M_{cc'}$, and $M_{vc}$, required for the $\textbf{k}$-point pairs it owns through serial FFTs. It then computes the sums in Eq. \ref{direct_term} and Eq. \ref{exchange_term} for all of its pairs.  

If the number of CPUs is greater than $N_k^2$, as is often the case for large systems and molecules, but less than $N_k^2 \cdot N_c^2$, we distribute the $(c{\bf k},c'{\bf k}')$ pairs evenly among the processors, first distributing the processors evenly over ${\bf k}$-point pairs and then creating pools to distribute the $(c,c')$ evenly among the pools. In this scheme, each CPU computes $M_{cc'}$ for only the $(c{\bf k},c'{\bf k}')$ it owns but computes $M_{vv'}$ and $M_{vc}$ for all $v$ and $v'$ at each $(c{\bf k},c'{\bf k}')$ pairs it owns. Each CPU does the summations in Eq. \ref{direct_term} and Eq. \ref{exchange_term} for all $v$, $v'$ for the $(c{\bf k},c'{\bf k}')$ it owns.

If the number of CPUs is greater than $N_k^2 \cdot N_c^2$, we distribute the entire set of $(vc{\bf k},v'c'{\bf k}')$ pairs out evenly among the processors, first distributing the processors evenly over $(c{\bf k},c'{\bf k}')$ pairs and then creating pools to distribute the $(v,v')$ evenly among the pools. In this scheme, each CPU computes only the $M_{vv'}$, $M_{cc'}$, and $M_{vc}$ for the $(vc{\bf k},v'c'{\bf k}')$ pairs it owns. Additionally, each CPU does the summations in Eq. \ref{direct_term} and Eq. \ref{exchange_term} for only the pairs it owns.  

In the last scheme the calculation of the matrix elements has a parallel wall-time scaling of $N$ and the summations scale as $N^2$ (accounting for the sum over ${\bf GG}'$) if a limitless number of CPU resources is assumed.

The large arrays that must be stored in memory are the dielectric matrix, the wavefunctions and the computed kernel itself.  The computed kernel is distributed evenly in memory among the processors and computed directly by the CPUs which own the various $(vc{\bf k},v'c'{\bf k}')$ pairs.  The dielectric matrix is distributed, as in the \texttt{sigma} executable, over its rows ${\bf G}$ and must be broadcast during each calculation of the sums in Eq. \ref{direct_term} and Eq. \ref{exchange_term}. It is for the purposes of minimizing the communication of the dielectric matrix that we use the three different parallelization schemes above -- \textit{i.e.} so that we might work on the biggest blocks of $(v,v')$ and $(c,c')$ at once. For example, in the first scheme, where the number of CPUs is less than $N_k^2$, we do the sums in Eq. \ref{direct_term} and Eq. \ref{exchange_term} for all $(vc,v'c')$ at once, so that we need only broadcast the dielectric matrix one time.

\begin{figure}
\begin{center}
\includegraphics[width=6.324cm] {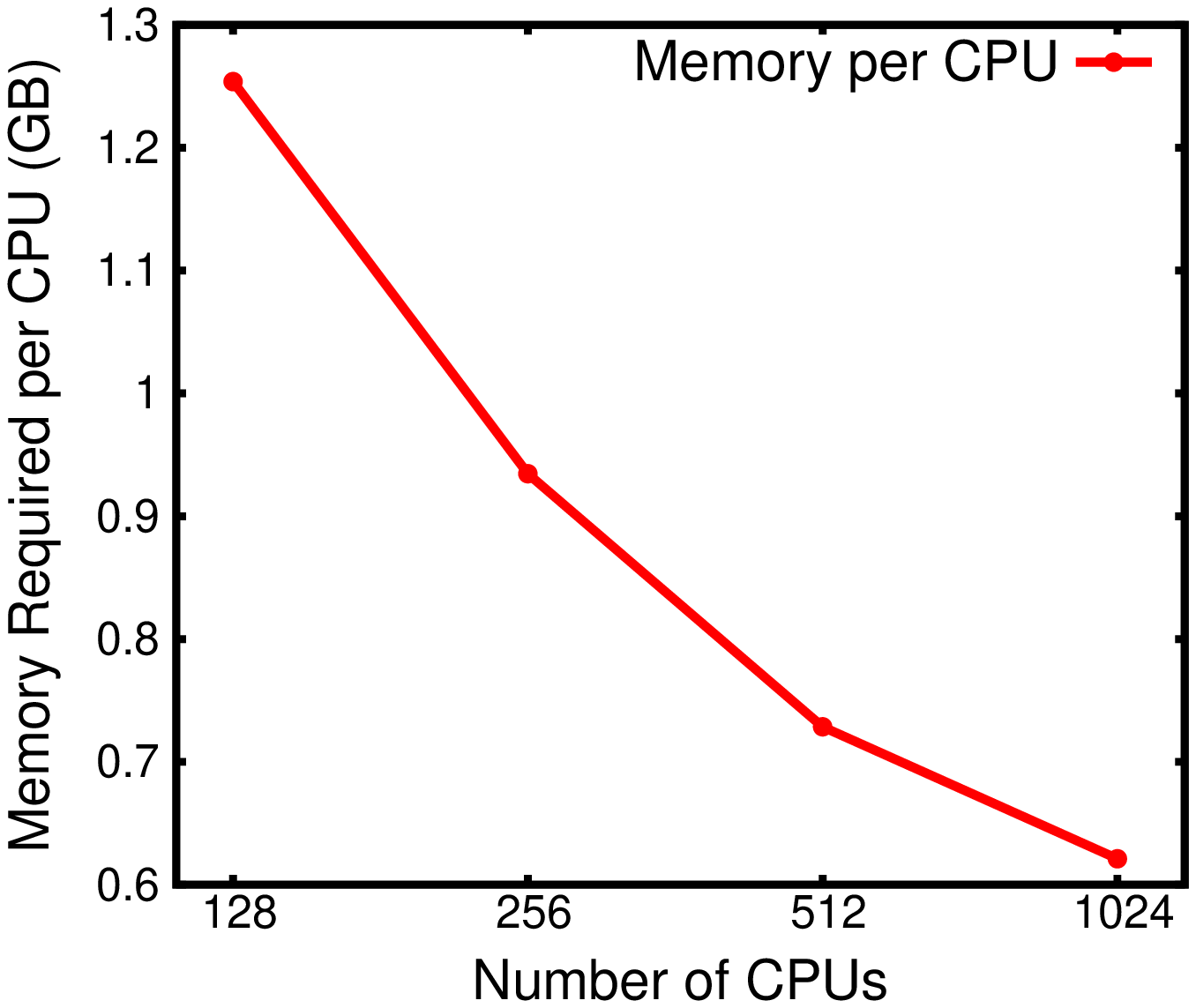}
\includegraphics[width=6.324cm] {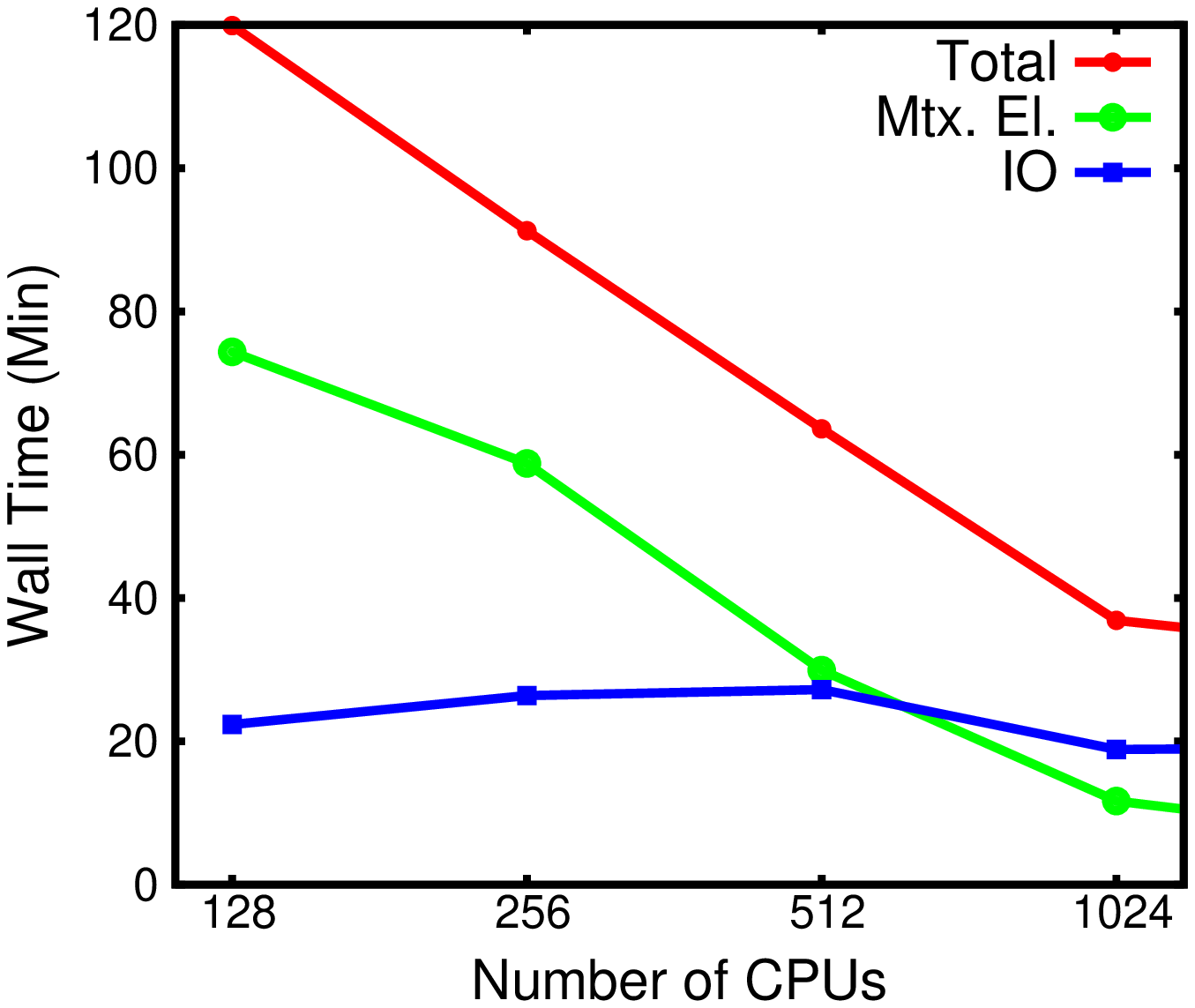}
\end{center}
\caption{(Left) Memory per CPU required vs. the number of CPUs for a \texttt{kernel} calculation on the (20,20) SWCNT. (Right) The wall-time required \textit{vs.} the number of CPUs used for a \texttt{kernel} calculation on the (20,20) SWCNT. The parameters used are described in the text.}
\label{xctTime}
\end{figure}

The wall-time scaling for the example \texttt{kernel} (20,20) SWCNT calculation is shown in Fig. \ref{xctTime}. We see nearly linear scaling up to 1024 CPUs -- which is the square of the number of {\bf k}-points, 32.

In the \texttt{absorption} executable, as described above, the first computational challenge is the interpolation of the kernel from the coarse $\textbf{k}$-grid onto the fine $\textbf{k}$-grid. The computation of the interpolation coefficients, Eq. \ref{coefficients}, is done by distributing the fine-grid $\textbf{k}$-points evenly among the processors. For molecules or other large systems, this does not represent a problem because the computation of the coefficients is very quick in these cases regardless of the lack of parallelization.

The parallelization of the kernel interpolation is different from the parallelization scheme in the \texttt{kernel} executable.  However, the parallelization is also described by three different schemes:

First, if the number of CPUs is less than $N_k$, where $N_k$ is the number of fine-grid ${\bf k}$-points, we distribute the $N_k$ {\bf k}-vectors on the fine grid evenly among the CPUs.  Each processor owns all the $(vc{\bf k},v'c'{\bf k}')$ pairs consistent with the ${\bf k}$-vectors it was assigned. It then performs the interpolation in Eq. \ref{interpolation} serially by replacing the simple loops that represent the sums with four matrix-matrix multiplications. For example, for each $n_1$ and $n_3$, one can write the sum over $n_2$ as a matrix-matrix product between the coarse kernel matrix (whose outer dimension is $n_4$) and the $\left[ C_{v,n_2}^{{\bf k}_{\text{co}}} \right]^{*}$ matrix (whose outer dimension is $N_v$ on the fine grid). We write the remaining sums as a similar matrix-matrix product.

If the number of CPUs is greater than $N_k$ but less than $N_k \cdot N_c$, we distribute the $N_k \cdot N_c$ ${\bf k}$-point and conduction-band pairs evenly among the processors. Again, each processor owns all the $(vc{\bf k},v'c'{\bf k}')$ pairs consistent with the $({\bf k},c)$ it was assigned. In the previous scheme, we utilized matrix-matrix products to interpolate many kernel elements at once. This required the allocation of an array of size $N_v^2 \cdot N_c^2$ as described above.  In the present scheme, we avoid storing large intermediate arrays by doing more of the sums in Eq. \ref{interpolation} as simple loops rather than matrix products. By default we do two simple loops of two matrix products. However, if the user selects the \texttt{low\_memory} option, we do all four summations as four nested loops without the aid of matrix-matrix multiplications, obtaining one fine-grid matrix element at a time without the need for any intermediate matrices.

If the number of CPUs is greater than $N_k \cdot N_c$, we distribute the $N_k \cdot N_c \cdot N_v$ $\textbf{k}$-point and conduction- and valence-band pairs evenly among the processors. Each processor owns all the $(vc{\bf k},v'c'{\bf k}')$ pairs consistent with the $({\bf k},c,v)$ it was assigned. The interpolation is done exactly as in the previous case.

Once the fine-grid kernel has been constructed, in the \texttt{absorption} executable, the matrix is diagonalized using ScaLAPACK with a block-cyclic layout \cite{scalapack}. This diagonalization scales well to $O(1000)$ CPUs but saturates quickly beyond this point. 
In order to calculate the absorption spectrum as per Eq. \ref{eps2}, we compute all the necessary matrix elements, Eq. \ref{velocityop}, by distributing the fine-grid $\textbf{k}$-points evenly over processors.
In order to diagonalize large matrices ($e.g.$, graphene on a $256 \times 256$ ${\bf k}$-point grid \cite{yang09}), we turn to iterative methods, in particular Haydock recursion \cite{haydock80,benedict99}. Since it requires only matrix-vector products, this method scales well to larger numbers of processors. It should be pointed out that the kernel matrix is not sparse in general, so methods designed for the diagonalization of sparse matrices are not appropriate here.

%=======================================================================================
%=======================================================================================

\section{Coulomb Interaction}
\label{sec:truncation}

The bare Coulomb interaction is used in many places throughout the code. In all cases, the 1D array $v({\bf q}+{\bf G})$ is generated from a single \texttt{vcoul\_generator} routine in the \texttt{Common} directory.  However, there is a lot that can be specified about the Coulomb interaction in the code. The Coulomb interaction can be truncated to eliminate the spurious interaction between periodic images of nano-systems in a super-cell calculation. One can implement a cell-averaging technique whereby the value of the interaction at each ${\bf q}$-point (or ${\bf q}\rightarrow0$ in particular) can be replaced by the average of $v({\bf q}+{\bf G})$ in the volume the ${\bf q}$-point represents. Finally, this average can be made to include the ${\bf q}$-dependence of the inverse dielectric function also if $W$ is the final quantity of relevance for the application -- such as in the evaluation of $W$ for the self energy.

The \texttt{BerkeleyGW} package contains five general choices for the Coulomb interaction. Firstly, one may choose to use the bulk, untruncated value expressed in Eq. \ref{untruncated_bare_coulomb}.  There are in addition 4 choices of Coulomb interaction that truncate the interaction beyond a certain cutoff in real space, of generic form
\begin{equation}
v_{\rm t}({\bf r})=\frac{\Theta(f({\bf r}))}{r}
\end{equation}
where $f$ is some function that describes the geometry in which the interaction is truncated. The four choices available implement the methods of Ismail-Beigi \cite{beigi06}: Wigner-Seitz slab truncation, Wigner-Seitz wire truncation, Wigner-Seitz box truncation, and spherical truncation.  The Wigner-Seitz box truncation and the spherical truncation truncate the Coulomb interaction in all three spatial directions, yielding a finite value of $v_{\rm t}({\bf q}=0)$.  All the Wigner-Seitz truncation schemes truncate the interaction at the edges of the Wigner-Seitz cell in the non-periodic directions.  Slab truncation is intended for nano-systems with slab-like geometry. The Coulomb interaction is truncated at the edges of the unit cell in the direction ($z$) perpendicular to the slab plane ($xy$).  Wire truncation is intended for nano-systems with wire-like geometry. The Coulomb interaction is truncated at the edges of the unit cell in the two directions ($xy$) perpendicular to the wire axis ($z$). Spherical truncation allows the user to specify manually a spherical truncation radius outside of which the Coulomb interaction will be truncated.

Like the untruncated interaction, the slab-truncation and spherical-truncation schemes have the benefit that $v_{\rm t}({\bf q}+{\bf G})$ can be constructed analytically:
\begin{equation}
v^{\rm sph}_{\rm t}({\bf q})=\frac{4\pi}{q^2}\cdot(1-\cos(r_{\rm c}\cdot q))
\end{equation}
\begin{equation}
v^{\rm slab}_{\rm t}({\bf q})=\frac{4\pi}{q^2}\cdot(1-e^{-q_{xy}\cdot z_c}\cos(q_z \cdot z_{\rm c}))
\end{equation}
where $r_{\rm c}$ and $z_{\rm c}$ are the truncation distances in the radial and perpendicular directions, respectively. The wire-truncation and box-truncation on the other hand are computed numerically through the use of FFTs.  First, the truncated interaction, ($2K_0(|q_z|\rho)$ for wire geometry where $\rho=\sqrt{x^2+y^2}$ and $K_0$ is the modified Bessel function and $1/r$ for box geometry), is constructed on a real-space grid in the Wigner-Seitz cell and folded into the traditional unit cell. The real-space grid is typically more dense than the charge density grid used in DFT calculations. The density of points on the real-space grid relative to the charge density grid in $xy$ directions for wire-truncation and in $xyz$ directions for box-truncation is set to be a factor of 4 and 2, respectively. The origin of the Coulomb potential is offset from the origin of the coordinate system by half a grid step to avoid the singularity. We then FFT to yield the $v_{\rm t}({\bf q})$ directly. The FFT is done in parallel. For wire-truncation, $xy$-planes are evenly distributed among processors and each processor performs 2D-FFTs in $xy$-planes it owns. For box-truncation, the parallel 3D-FFT is performed as follows: $xy$-planes are distributed and 2D-FFTs in $xy$-planes are carried out the same way as for wire-truncation; the data is then transferred from $xy$-planes to $z$-rods which also are evenly distributed among processors; finally, each processor performs 1D-FFTs in $z$-rods it owns. After the FFT, the origin of the Coulomb potential is shifted back to the origin of the coordinate system by multiplying $v_{\rm t}({\bf G})$ with $\exp(i2\pi{\bf G}\cdot\frac{1}{2}{\bf d})$ where ${\bf d}$ is the real-space grid displacement vector.

As mentioned above, for all interaction choices with the exception of cell-box and spherical truncation, the Coulomb interaction diverges as ${\bf q}\rightarrow0$. For the case of no truncation, $v({\bf q}\rightarrow0)\propto 1/q^2$; for slab truncation, $v^{\rm slab}_{\rm t}({\bf q}\rightarrow0)\propto 1/q$; for wire truncation, $v^{\rm wire}_{\rm t}({\bf q}\rightarrow0)\propto-\ln(q)$. As we mentioned in Sec. \ref{sec:xi0}, this divergence is handled in \texttt{epsilon} by taking a numerical limit -- that is, evaluating $\epsilon$ at a small but finite ${\bf q}_0$. For \texttt{sigma} and \texttt{absorption}, on the other hand, we are interested in evaluating directly $W_{{\bf GG}'}({\bf q})$ matrix elements and the appropriate treatment is to replace the divergent (for non-metals) $W({\bf q}\rightarrow0)$ with an average over the volume in reciprocal space that ${\bf q}=0$ represents:
\begin{equation}
W^{\rm avg}_{{\bf GG}'}({\bf q}=0)=\frac{N_{\bf q} \cdot \text{Vol}}{2\pi} \int_{\rm cell} d^n q\ W_{{\bf GG}'}({\bf q})
\label{cellaverage}
\end{equation}
where ``cell'' represents the volume in reciprocal space closer to ${\bf q}=0$ than any other ${\bf q}$-points, and $d^n q$ represents the appropriate dimensional differential for the truncation scheme (e.g., $d^2 q$ for slab truncation).  Eq. \ref{cellaverage} yields a finite number in all truncation schemes even when $W({\bf q}\rightarrow0)$ itself is divergent because of the reduced phase-space around ${\bf q}\rightarrow0$. 

For metallic systems, it is particularly important to use Eq. \ref{cellaverage} to average $W$, as opposed to averaging $v({\bf q})$ and $\epsilon^{-1}({\bf q})$ separately, since for metals $\epsilon^{-1}({\bf q})$ has inverse ${\bf q}$-dependence from $v({\bf q})$ at small ${\bf q}$, yielding a constant limit: $W^{\rm metal}({\bf q}\rightarrow0)=C$. The user can tell the code which model to use for the ${\bf q}$-dependence of $\epsilon^{-1}$ by specifying one of the screening flags in the input files. The ${\bf q}\rightarrow{\bf 0}$ limits for the inverse dielectric function and screened Coulomb interaction are enumerated in Table \ref{w_table} for the semiconductor and metallic screening types.

\begin{figure}
\begin{center}
\includegraphics[width=6.324cm] {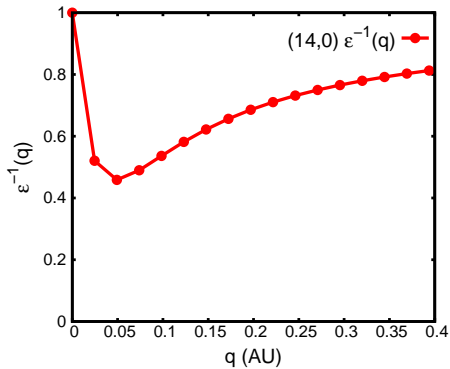}
\end{center}
\caption{$\epsilon^{-1}(q)$ in the (14,0) single-walled carbon nanotube for ${\bf q}$ along the tube axis as reported in the \texttt{epsilon.log} output file. The circles represent ${\bf q}$-grid points included in a $1 \times 1 \times 32$ sampling of the first Brillouin zone.}
\label{14-0}
\end{figure}

As shown in Fig. \ref{14-0}, including $\epsilon^{-1}$ in the cell-averaging scheme also is important when using a truncated interaction where $\epsilon({\bf q}=0) = 1$ but quickly rises by the first non-zero ${\bf q}$-point \cite{beigi06}. The figure shows $\epsilon^{-1}(q)$ along the tube axis in the (14,0) SWCNT. $\epsilon^{-1}(0)=1$ but decreases nearly by half by the first non-zero grid point included in a $1 \times 1 \times 32$ sampling of the first Brillouin zone. $\epsilon^{-1}(0)=1$ is a general property of truncated systems with semiconductor-type screening since the $q^2$ dependence of the polarizability approaches $0$ faster than the Coulomb interaction diverges at $q\rightarrow0$. \texttt{BerkeleyGW} uses this $W$-averaging procedure by default for cases with truncated Coulomb interaction.

In general, using an extension of Eq. \ref{cellaverage} even for ${\bf q},{\bf GG}' \neq 0$ can speed up the convergence of a \texttt{sigma} or \texttt{absorption} calculation with respect to the number of ${\bf q}$-points required in the calculation. This can be explained easily by the fact that one is replacing a finite sum over ${\bf q}$-points with an integral -- mimicking a calculation on a much larger set of ${\bf q}$-points or a much larger unit cell. The user can ask that \texttt{BerkeleyGW} use the cell-averaged $W$ for all ${\bf q}$ and ${\bf G}$ below an energy cutoff specified in the input file.

\begin{table}
\begin{center}
\begin{tabular}{ | c | c | c | c | c | }  \hline 
$\epsilon_{{\bf GG}'}^{-1}$ & Semiconductor & Semiconductor & Metal & Metal \\ 
& & truncation & & truncation \\ \hline & & & & \\ 
Head & Constant & $1$ &  $q^2$ & $\frac{1}{v_{\rm t}(q)}$ \\
[.2in] Wing & $q$ & $q\epsilon^{-1}_{00}\left(q\right)$ & $q^2$ & $\frac{1}{v_{\rm t}(q)}$ \\
[.2in] Wing$'$ & $\frac{1}{q}$ & $q\epsilon^{-1}_{00}\left(q\right)v_{\rm t}(q)$ & Constant & Constant \\[.2in] \hline
\end{tabular}

\begin{tabular}{ | c | c | c |  c | c | } \hline 
$W_{{\bf GG}'}$ & Semiconductor & Semiconductor & Metal & Metal \\ 
& & truncation & & truncation \\ \hline & & & & \\ 
Head & $\frac{1}{q^2}$ & $\epsilon^{-1}_{00}\left(q\right)v_{\rm t}(q)$ & Constant & Constant \\
[.2in] Wing & $\frac{1}{q}$ & $q\epsilon^{-1}_{00}\left(q\right)v_{\rm t}(q)$ & Constant & Constant \\
[.2in] Wing$'$ & $\frac{1}{q}$ & $q\epsilon^{-1}_{00}\left(q\right)v_{\rm t}(q)$ & Constant & Constant \\[.2in] \hline 
\end{tabular}
\end{center}
\caption{Top: ${\bf q}\rightarrow{\bf 0}$ limits of the head $\epsilon_{{\bf 00}}^{-1}({\bf q})$, wing $\epsilon_{{\bf G0}}^{-1}({\bf q})$, and wing$'$ $\epsilon_{{\bf 0G}'}^{-1}({\bf q})$, of the inverse dielectric matrix. Bottom: ${\bf q}\rightarrow{\bf 0}$ limits of the head and wings of the screened Coulomb interaction, $W_{{\bf GG}'}({\bf q})$.}
\label{w_table}
\end{table}

The averaging is implemented in the code using a Monte Carlo integration method with 2,500,000 random points in each cell.

\subsection{Grid Uniformity}

Another important consideration in performing integrals over ${\bf q}$ of the Coulomb interaction in \texttt{sigma} and \texttt{kernel} is the sampling of the grid on which the integral is done. If the sampling in different directions is very different, then the result of the integrals will not go to the correct limit, since it will resemble a 1D or 2D integration rather than a 3D integration. This issue is important for calculation of nano-systems without truncation, or for highly anisotropic crystals. We determine the effective sampling in each direction as follows: take the vectors $b_i / N_i$, where $b_i$ is a reciprocal lattice vector and $N_i$ is the number of ${\bf q}$-points in that direction. Find the shortest vector. Orthogonalize the next shortest vector to that one. Orthogonalize the remaining vector to the first two. (It is important to use this order since orthogonalization always makes the vector shorter.) Now compare the lengths of these orthogonalized vectors: if the ratio between the longest and shortest is too large (we use a factor of 2 as a tolerance), the grid is non-uniform and may give incorrect answers. The code will write a warning in this case, and the user should try to use a more nearly uniform grid, or check the convergence of results against the cell-averaging cutoff. Note that the sampling in any direction in which the Coulomb interaction is truncated is irrelevant when checking for grid uniformity.

%========================================================
%========================================================

\section{Symmetry and degeneracy}

\subsection{Mean field}
\label{sec:symmetry_mf}

As was mentioned in the Introduction, the largest cost when performing a GW calculation with the \texttt{BerkeleyGW} package is the generation of the input mean-field states. In order to reduce this cost, all the codes allow the user to input the wavefunctions in only the reduced Brillouin zone and construct the wavefunctions in the full zone by the following relation:
\begin{equation}
\phi_{{\bf R}({\bf k})}({\bf G})=\phi_{\bf k}({\bf R}^{-1}{\bf (G)}) e^{-i {\bf G} \cdot {\bf \tau}}
\label{wfnsymmetry}
\end{equation}
where the symmetry operation is defined by a reciprocal-space rotation matrix ${\bf R}$ and a fractional translation ${\bf \tau}$ such that ${\bf x}' = {\bf R}^{-1} {\bf x} + {\bf \tau}$.

\begin{figure}
\begin{center}
\includegraphics[width=2cm] {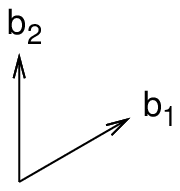} \includegraphics[width=5cm] {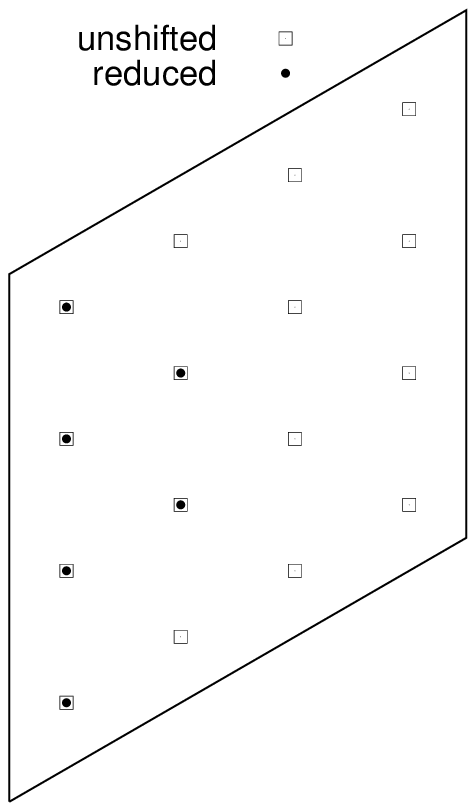} \includegraphics[width=5cm] {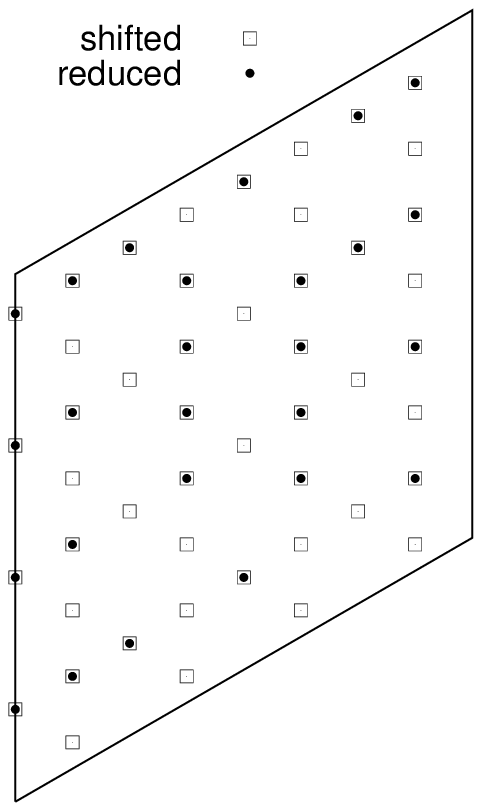}
\end{center}
\caption{An example of the construction of 4 $\times$ 4 main and shifted ${\bf k}$-grids for graphene. (Left) The main grid has a (0.5, 0.5) shift (crystal coordinates). There are 16 points in the full Brillouin zone and 6 irreducible points. (Right) The shifted grid has a (0.0, 0.05) shift (crystal coordinates). There are 48 points in the full Brillouin zone and 26 irreducible points.}
\label{kgrid}
\end{figure}

The main ${\bf k}$-grid used is generated by constructing a uniform grid (with a possible shift, typically half a unit as in Monkhorst-Pack grids) and then reducing by the symmetry operations. The shifted grid used in \texttt{epsilon} for constructing \texttt{eps0mat} at a small ${\bf q}$-vector is generated by unfolding this reduced set of points with all the symmetry operations, reducing again by the symmetry operations of the subgroup of the ${\bf q}$-vector, and then applying a small ${\bf q}$-shift. (Note that, contrary to a na\"{i}ve expectation, this reduced shifted grid may contain many more points than the original uniform grid.) An example for graphene is given in Fig. \ref{kgrid}. This procedure provides all the ${\bf k}$-points needed for calculations in \texttt{BerkeleyGW}, and it is implemented in the utility \texttt{kgrid.x} in the \texttt{MeanField/ESPRESSO} directory, which should be used to generate the set of ${\bf k}$-points in mean-field calculations. \texttt{PARATEC} also has built-in support for this construction.
The symmetry analysis in \texttt{kgrid.x}, \texttt{epm2bgw.x}, and \texttt{siesta2bgw.x} is performed with the spglib library \cite{spglib}.

\subsection{Dielectric matrix}

The wavefunctions in the full zone then are used for the sum over ${\bf k}$ in Eq. \ref{eqn_static_xi} in \texttt{epsilon}, \texttt{sigma}, \texttt{kernel} and \texttt{absorption}, which requires not only the wavefunctions in the full zone but also the dielectric matrix. While in principle one could construct the dielectric matrices at all the {\bf q}-points required in Eqs. \ref{ff-sx} -- \ref{hf-x} and \ref{direct_term}, in practice one can use symmetry to reduce the required {\bf q}-points. The \texttt{epsilon} code requires the user to calculate the dielectric matrices on a reduced set of {\bf q}-points and the other codes generate the dielectric matrices in the full zone. If one defines ${\bf q}_1 = {\bf R}({\bf q}) + {\bf G}_{R}$, where ${\bf G}_{R}$ is a ${\bf G}$-vector chosen to ensure that ${\bf q}$ and ${\bf q}_1$ are in the first Brillouin zone, and ${\bf R}, {\bf \tau}$ are rotation matrix and translation respectively, then one can use the relation \cite{hybertsen86,hybertsen87} : 
\begin{equation}
\label{eqpsymmetry}
\epsilon^{-1}_{{\bf G}{\bf G}'}({\bf q}_1; E) = e^{-i ({\bf G} - {\bf G}') \cdot {\bf \tau}} \epsilon^{-1}_{{\bf G}_1{\bf G}'_1}({\bf q}; E)
\end{equation}
where ${\bf G}_1 = {\bf R}^{-1}({\bf G} + {\bf G}_{R})$. Given $\epsilon^{-1}$ in the reduced zone, this allows one to construct it in the full zone. 

\subsection{Truncation of sums}

Both the dielectric matrix and the self-energy operator involve infinite sums over unoccupied states, which must in practice be truncated in the \texttt{epsilon} and \texttt{sigma} codes. The dielectric matrix and self-energy operator only retain the full symmetry of the system if the truncation does not cut through any degenerate subspaces. Consider a subspace of states belonging to a degenerate representation of a symmetry operation. Only the whole subspace is invariant under that operation, while just a part of it is not necessarily. As a result, a calculation using only part of the subspace will produce self energies that break degeneracies due to that operation. Moreover, the actual values obtained are not well defined because the states used are arbitrary linear combinations in the subspace, which could even differ from run to run of a DFT code depending on how the calculation is initialized. These considerations are particularly acute for sums over a small number of states, since the contribution of the last few bands may be significant. Therefore, the \texttt{epsilon} and \texttt{sigma} codes check that the highest band requested for the sum is not degenerate with the next one, and block calculations that will break degeneracy. However, it is also possible to override this behavior with the flag \texttt{degeneracy\_check\_override}, for testing purposes and because in some cases there may be overlapping degenerate subspaces on different ${\bf k}$-points that make it difficult to find acceptable numbers of bands; for large numbers of bands, the effect of truncation in a degenerate subspace will be small.

\subsection{Self-energy operator}

Degeneracy is also important from the point of view of the states on which self energies are calculated, as opposed to those appearing in the sum. Since the self-energy operator has the full symmetry of the system, the matrix elements between states belonging to different representations are zero by symmetry. In the presence of high symmetry, this consideration can make the matrix quite sparse. To take advantage fully of symmetry here would require a careful analysis of each wavefunction's behavior under various symmetry operations and comparison to character tables of space groups. Users certainly can do this in deciding which off-diagonal self-energy matrix elements to calculate. The \texttt{Sigma} takes a very simple approach to identify some of the elements which are zero by symmetry, based on degeneracy. The multiplicity of the degenerate subspace to which each states belongs is counted (1, 2, or 3 for the standard space groups), and clearly two states in subspaces of different multiplicity must belong to different representations, and their matrix element can be set to zero without calculation. This saves time and enforces symmetry.

Application of symmetry in a degenerate subspace can also speed up calculation of diagonal elements of the self-energy operator. The expressions for the exchange, screened exchange, and Coulomb-hole parts contain a sum over ${\bf q}$. In general, this must be done over the whole Brillouin zone, but to calculate the sum of the self energies within a degenerate subspace it is sufficient to use the irreducible part of the Brillouin zone. Each part of $\Sigma$, in the various approximations, has the generic form

%\begin{widetext}
\begin{align}
\label{generic-sigma}
\left<n{\bf k}\right|\Sigma \left|n'{\bf k}\right>&=
-\sum_{n''} \sum_{{\bf qGG}'}
\left< n {\bf k} \left| e^{i \left( {\bf q} + {\bf G} \right) \cdot {\bf r}} \right| n'' {\bf k - q} \right> 
\left< n'' {\bf k - q} \left| e^{-i \left( {\bf q} + {\bf G'} \right) \cdot {\bf r}} \right| n' {\bf k} \right>
\\
\nonumber
&\times F \left({\bf q},{\bf G},{\bf G'}\right)
\end{align}
%\end{widetext}

The summand is invariant under application of a symmetry operation $O$ in the subgroup of ${\bf k}$ provided that $n = n'$ and $n$ and $n''$ are non-degenerate, since in that case the action of the operation simply introduces a phase: $O \left| m k \right> = e^{i \theta} \left| m k \right>$ (degenerate states may instead transform into linear combinations in the degenerate subspace). These phases are canceled by the fact that each state appears also with its complex conjugate. If the states $n''$ in the sum are degenerate, the summand is not invariant but the sum is, if the whole degenerate subspace is summed over, since then we are taking the trace of the projector matrix $\left| n'' {\bf k} \right> \left< n'' {\bf k} \right|$ in that subspace, which is invariant \cite{hybertsen87}. If $n$ is degenerate, then $\left<n{\bf k}\right|\Sigma \left|n{\bf k}\right>$ is not invariant, but the trace of the self-energy in the degenerate subspace, $\sum_n \left<n{\bf k}\right|\Sigma \left|n{\bf k}\right>$, is invariant. Therefore, to calculate diagonal elements for a whole degenerate subspace, for each state we sum only over ${\bf q}$ in the irreducible zone, with weight $W_{\bf q}$ from the number of ${\bf q}$-vectors related to ${\bf q}$ by symmetry. We then symmetrize by assigning the average to each:

%\begin{widetext}
\begin{align}
\label{generic-sigma2}
\left<m{\bf k}\right|\Sigma \left|m{\bf k}\right>&=\frac{1}{N_{\rm deg}} \sum_n^{\rm deg} \left<n{\bf k}\right|\Sigma \left|n{\bf k}\right> \nonumber & \\
& = -\sum_n^{\rm deg} \sum_{n''} \sum_{\bf GG'} \sum_{\bf q}^{\rm irr} W_{\bf q} 
\left< n {\bf k} \left| e^{i \left( {\bf q} + {\bf G} \right) \cdot {\bf r}} \right| n'' {\bf k - q} \right> 
\\
\nonumber
&\times
\left< n'' {\bf k - q} \left| e^{-i \left( {\bf q} + {\bf G'} \right) \cdot {\bf r}} \right| n {\bf k} \right>
F \left({\bf q},{\bf G},{\bf G'}\right)
\end{align}
%\end{widetext}

(This averaging over degenerate bands is also done to enforce symmetry even when we use the full ${\bf q}$-sum, since the results may differ slightly due to limited precision in the wavefunctions from the mean-field calculation.) If we are calculating only part of a degenerate subspace, this trick does not work, and we must perform the complete sum. For diagonal elements, the code by default uses the irreducible ${\bf q}$-sum and will write an error if the calculation requires the full sum because of degeneracy, directing the user to enable it via the flag \texttt{no\_symmetries\_q\_grid}, or include all states in the degenerate subspace. For off-diagonal elements ($n \ne n''$), even if both are non-degenerate, application of the symmetry operation, in general, introduces different phases from the two states, which are not canceled. Thus the contributions from different ${\bf q}$-points related by symmetry differ, so that the full sum must always be used.

\subsection{Bethe-Salpeter equation}

Degeneracy must be considered in BSE calculations as well, when choosing the subspace in which to work. If the set of occupied or unoccupied states includes only part of a degenerate subspace, then the solutions found by \texttt{absorption} will break symmetry and can give qualitatively incorrect results. For example, an excitation that should have zero oscillator strength by symmetry, due to interference between transitions to two degenerate states, may not be dark if only one of those transitions is included. This issue is quite general and applies to the choice of active spaces in other theories as well, such as configuration interaction \cite{gpzhang}. Breaking degeneracy in either the coarse or fine grid can also cause trouble in interpolation of the kernel and quasiparticle energies.

\subsection{Degeneracy utility}

We provide a utility called \texttt{degeneracy\_check.x} which reads wavefunction files and writes out a list of acceptable numbers of bands. Multiple wavefunction files can be checked at once, for example the shifted and unshifted grids in \texttt{Epsilon} or shifted, unshifted, coarse, and fine grids for Bethe-Salpeter equation calculations, in which case the utility will identify numbers of bands which are consistent with degeneracy for every file.

\subsection{Real and complex flavors}

The component executables come in two ``flavors,'' real and complex, specified at compile time and denoted by the suffix \texttt{.real.x} or \texttt{.cplx.x}. When the system has inversion and time-reversal symmetry, we can choose the wavefunctions to be real in reciprocal space. The plane-wave expansions are:
\begin{align}
u \left( \bf{r} \right) = \sum_{\bf{G}} u_{\bf{G}} e^{i \bf{G} \cdot \bf{r}} \\
u \left( \bf{-r} \right) = \sum_{\bf{G}} u_{\bf{G}} e^{-i \bf{G} \cdot \bf{r}} \\
u^{*} \left( \bf{r} \right) = \sum_{\bf{G}} u^{*}_{\bf{G}} e^{-i \bf{G} \cdot \bf{r}}
\end{align}
The symmetry conditions mean that wavefunctions can be chosen to satisfy
$u \left( -{\bf r} \right) = a u \left( {\bf r} \right)$ (inversion symmetry) and
$u^{*} \left( {\bf r} \right) = b u \left( {\bf r} \right)$
(time-reversal, equivalent to taking the complex conjugate of the Schr\"odinger equation),
with $a, b$ each equal to $\pm 1$ depending on whether the wavefunction belongs to an odd or even representation.
Thus we can choose $u \left( -{\bf r} \right) = c u^{*} \left( {\bf r} \right)$ with $c = a b$ also equal to $\pm 1$.
Combining this with the plane-wave expansions,
\begin{align}
\sum_{\bf{G}} u_{\bf{G}} e^{-i \bf{G} \cdot \bf{r}} = c \sum_{\bf{G}} u^{*}_{\bf{G}} e^{-i \bf{G} \cdot \bf{r}} \\
u_{\bf{G}} = c u^{*}_{\bf{G}}
\end{align}
The choice $c = 1$ corresponds to real coefficients; $c = -1$ corresponds to pure imaginary coefficients. Most plane-wave electronic-structure codes always use complex coefficients, and so the coefficients will in general not be real, even in the presence of inversion and time-reversal symmetry. For a non-degenerate state, the coefficients will be real times an arbitrary global phase, determined by the initialization of the solution procedure. We must divide out this global phase to make the coefficients real. In a degenerate subspace, the states need not be eigenstates of inversion, and so in general they may not just be real times a global phase. Instead, in each subspace of degeneracy $n$ we take the 2$n$ vectors given by the real and imaginary parts of each wavefunction, and then use a Gram-Schmidt process to find $n$ real orthonormal wavefunctions spanning the subspace.

The density and exchange-correlation potential are real already in the presence of inversion symmetry and there is no arbitrary phase possible.
The real-space density is always real: $\rho \left( {\bf r} \right) = \rho^{*} \left( {\bf r} \right)$.
With inversion symmetry, we also have $\rho \left( {\bf r} \right) = \rho \left( -{\bf r} \right)$.
In reciprocal space,
\begin{align}
\rho \left( {\bf r} \right) = \sum_{\bf G} \rho_{\bf G} e^{i {\bf G} \cdot {\bf r}} \\
\rho^{*} \left( {\bf r} \right) = \sum_{\bf G} \rho^{*}_{\bf G} e^{-i {\bf G} \cdot {\bf r}} \\
\rho \left( -{\bf r} \right) = \sum_{\bf G} \rho_{\bf G} e^{-i {\bf G} \cdot {\bf r}}
\end{align}
Together, these relations imply $\rho_{\bf G} = \rho^{*}_{\bf G}$, \textit{i.e.} the reciprocal-space coefficients are real. Precisely the same equations apply for the exchange-correlation potential.

The wavefunction, density, and exchange-correlation potential are then all stored as real coefficients, saving disk space (for the files), memory, and operations compared to the complex representation. Usually, only the lack of inversion symmetry of the lattice and basis would require the use of complex wavefunctions, but if spin-orbit coupling or magnetic fields are present, then time-reversal symmetry is lost and complex wavefunctions again are required.

\section{Computational Issues}

\subsection{Memory estimation}

In the beginning of each run, all the major code components print the amount of memory available per CPU and an estimate of memory required per CPU to perform the calculation. If the latter exceeds the former, the job is likely to fail with a memory allocation error. The amount of memory required is estimated by determining the sizes of the largest arrays after reading in the parameters of the system from the input files.
A straightforward approach to estimating the amount of available memory is to allocate memory by incremental amounts until the allocation call returns with an error. Unfortunately, in many implementations the allocation call returns without an error even if the requested amount of memory is not physically available, but the system fails when trying to access this ``allocated'' memory.
We implement another approach based on the Linux \texttt{/proc} file system. First, each CPU opens file \texttt{/proc/meminfo} and reads in the values of \texttt{MemFree} and \texttt{Cached}. The sum of these two values gives the amount of memory available per node.  This approach works on almost all modern high-performance computing systems where the Linux \texttt{/proc} File System is accessible. (However, for BSD-based MacOS which lacks \texttt{/proc/meminfo}, we read the page size and number of free and speculative pages from the command \texttt{vm\_stat}.)
Second, each CPU calls an intrinsic Fortran routine that returns the host name which is unique for each node. By comparing host names reported by different CPUs we identify the number of CPUs per node. The amount of memory available per CPU is then given by the ratio of the amount of memory available per node to the number of CPUs per node.

\subsection{Makefiles}

The main codes are in the \texttt{Epsilon}, \texttt{Sigma}, \texttt{BSE}, \texttt{PlotXct}, and \texttt{MeanField} directories. Routines used by all parts are in the \texttt{Common} directory, and routines common to some of the \texttt{MeanField} codes are in the \texttt{Symmetry} directory. The \texttt{Makefile}s are designed for GNU Make, and enable targets in a directory to be built from any level of the directory hierarchy. They contain a full set of dependencies, including those between directories, to ensure that the build is correct after any changes to source, for ease in development and modification. This also enables use of parallel make on large numbers of processors for rapid builds -- any omissions in the dependencies generally cause a failure for a parallel \texttt{make}. The special \texttt{make} target \texttt{all-j} (\textit{i.e.} \texttt{make -j all-j}) begins by using all processes to build common directories, which contain files required by files in a large number of directories; otherwise, the build would fail due to attempts by multiple processes to read and write the same files in these directories.
Commonly, Fortran Makefiles are set up with object files depending on other object files. However, the real situation is that object files depend on module files (\texttt{.mod}) for the modules they use, and only executables depend on object files. Therefore we have dependencies directly on the module files to ensure the required files are present for compilation, particularly for parallel builds.
%Unfortunately \texttt{gfortran} and \texttt{g95} will not update module files upon compilation if the interfaces have not changed, which causes these files to be recompiled unecessarily. We remedy this behavior by deleting the module files before compilation when using these compilers.

\subsection{Installation instructions}

The code can be installed via the following steps:
\begin{quote}
\begin{verbatim}
cp [flavor_real.mk/flavor_cplx.mk] flavor.mk 
ln -s config/[mysystem].mk arch.mk
make all
make check[-jobscript]
\end{verbatim}
\end{quote}

First a flavor is selected by copying the appropriate file to \texttt{flavor.mk}. Then a configuration file must be put as \texttt{arch.mk}. Configurations appropriate for various supercomputers as well as for using standard Ubuntu or Macports packages are provided in the \texttt{config} directory. Appropriate paths, libraries, and compiler flags can be set in a new \texttt{arch.mk} for other systems. Finally the test suite should be run to confirm that the build is working. In serial, the command is \texttt{make check}; for parallel builds, it is \texttt{make check-parallel}. On machines with a scheduler, a job script should be created to run this command. For the architectures supported in \texttt{config}, job scripts to run the test suite are provided in the \texttt{testsuite} directory, and can be used via \texttt{make check-jobscript}.

\subsection{Validation and verification}

The importance of verification and validation of complicated scientific software packages is receiving increasing attention.
We use standard open-source tools for code development, following accepted best practices \cite{sciprog}. Development is done with the subversion (SVN) version-control system \cite{SVN} and Trac, an issue-tracking system and interface to SVN \cite{Trac}. All code runs identify the version and revision number used in the output for traceability of results, implemented via a special source file called \texttt{svninfo.f90} which all SVN revisions must modify (enforced via a pre-commit hook). Debug mode can be enabled via \texttt{-DDEBUG} in the \texttt{arch.mk} file, which performs extra checking including of dynamic memory allocation and deallocation. A macro enables a check of the status returned by the system after an allocation attempt, and reports failures, identifying the array name, size, source file, and line number, as well as which processor failed to allocate the array. Additionally, it keeps track of the amount of memory dynamically allocated and deallocated, so the code can report at the end of each run how much memory remains allocated, and the maximum and minimum memory `high-water-mark' among the processors. In debug mode, a stack trace also can be enabled, either on just the root processor, or on all processors (causing the code to run much slower), which can be used to locate where problems such as segmentation faults are occurring (possibly on only one processor). Verbose mode can be enabled via \texttt{-DVERBOSE} which writes extra information as the calculation proceeds.

The package contains a comprehensive test suite to test the various executables, run modes, and options, in the \texttt{testsuite} directory. Calculations of several different physical systems, with mean-field, \texttt{epsilon}, \texttt{sigma}, and BSE calculations, are carried out (including use of \texttt{PlotXct} and some utilities), detecting any run-time errors and showing any warnings generated. Then selected results are extracted from the output and compared to reference information within a specified tolerance. The actual calculated values, as well as timing for each step, are displayed. Each match is shown as either \texttt{OK} or \texttt{FAIL}, and a final summary is written of failures. The calculations are small and generally underconverged, to make them quick enough for routine testing and rapid feedback. The mean-field steps are either EPM (quick serial calculations) or stored compressed output from DFT calculations. The \texttt{Epsilon}, \texttt{Sigma}, and BSE calculations are run either in serial or on 4 processors (for parallel builds).

The test suite has numerous uses. It is useful for users to verify the success of a new build of the code on their platform (failures could be due to library problems, excessive optimizations, etc.). It is used for developers to verify that the code is giving reproducible answers, ensure consistency between serial/parallel runs, as well as real/complex and spin-polarized/unpolarized runs, and check that the code works with new compilers or libraries. On a routine basis, the test suite is also useful for developers to check that changes to the code do not introduce problems.
The driver scripts (\texttt{run\_testsuite.sh} and \texttt{run\_regression\_test.pl}) and specifications for the files defining the test steps are originally based on, and developed in conjunction with, those of the Octopus code \cite{octopus_pssb, octopus_CPC}. This framework is quite general and can be used easily for constructing a test suite for another code. It can be run in serial with the command \texttt{make check} (or \texttt{make check-save} to retain the working directories from the runs), or in parallel with \texttt{make check-jobscript} (or \texttt{make check-jobscript-save}). The system configuration file \texttt{arch.mk} can specify how to submit an appropriate jobscript for parallel execution on a supercomputer using a scheduler. Scripts are provided in the \texttt{testsuite} directory for some supercomputers.
The test suite is used with a continuous-integration system, the open-source tool BuildBot \cite{BuildBot}, to ensure the integrity of the code during development. Each commit to the SVN repository triggers a build of the code on each of 10 ``buildslaves,'' which have different configurations with respect to serial/parallel, compilers, and libraries. After the build, the test suite is run. BuildBot will report to the developers if the either the build or test runs failed, so the problem can be quickly remedied. Use of the various different buildslave configurations helps ensure that the code remains portable across different platforms and in accordance with the language standards. Two of the buildslaves are on a supercomputer with a scheduler, a situation for which standard BuildBot usage is problematic. We provide a Perl script \texttt{buildbot\_pbs.pl} that can submit jobs, monitor their status, capture their output for BuildBot, and determine success or failure. This script is general for any PBS scheduler and can be used for other codes too.

\subsection{Supported operating systems, compilers, and libraries}

With the test suite, we have tested the code extensively with various configurations, and support the following compilers and libraries:
\begin{itemize}
\item Operating systems: Linux, AIX, MacOS
\item Fortran compilers (required): \texttt{pgf90, ifort, gfortran, g95, openf90, sunf90, pathf90, crayftn}, \texttt{af90} (Absoft), \texttt{nagfor}, \texttt{xlf90} (experimental)
\item C compilers (optional): \texttt{pgcc, icc, gcc, opencc, pathcc, craycc}
\item C++ compilers (optional): \texttt{pgCC, icc, g++, openCC, pathCC, crayCC}
\item MPI implementation (optional): OpenMPI, MPICH1, MPICH2, MVAPICH2
\item LAPACK/BLAS implementation (required): NetLib, ATLAS, Intel MKL, ACML, Cray LibSci
\item ScaLAPACK/BLACS implementation (required by BSE if MPI is used): NetLib, Cray LibSci, Intel MKL, AMD
%(not Ubuntu packages)
\item FFTW (required): versions 2.1.5, 2.1.5.1, 2.1.5.2
\end{itemize}

\section{Utilities}
\label{sec:utilities}

\subsection{Visualization}

Several visualization tools designed to simplify work with the DFT codes and the \texttt{BerkeleyGW} package are provided in directory \texttt{Visual}. These include the \texttt{surface.x} code and Matter library.

Surface is a C++ code for generating an isosurface of a volumetric scalar field (such as the wave function, charge density, or local potential). The scalar field is read from a Gaussian Cube or XCrySDen XSF file, the surface triangulation is performed using the marching-cubes or marching-tetrahedra algorithms, and the isosurface is written in the POV-Ray scripting language. The final image is rendered using the ray-tracing program POV-Ray \cite{povray}. Running the surface code requires a fairly complicated input parameter file. A brief description of the input file parameters is given in the header of \texttt{surface.cpp}.

Matter is a Python library for manipulating atomic structures with periodic boundary conditions. It can translate and rotate the atomic coordinates, generate super-cells, assemble atomic systems from fragments, and convert between different file formats. The supported file formats are \texttt{mat}, \texttt{paratec}, \texttt{vasp}, \texttt{espresso}, \texttt{siesta}, \texttt{tbpw}, \texttt{xyz}, \texttt{xsf}, \texttt{wien} and \texttt{povray}. \texttt{mat} is the native file format of the library. \texttt{paratec}, \texttt{vasp}, \texttt{espresso}, \texttt{siesta}, \texttt{wien} and \texttt{tbpw} represent the formats used by different plane-wave and local-orbital DFT codes and an empirical pseudopotential code. \texttt{xyz} is a simple format supported by many molecular viewers, and \texttt{xsf} is the internal format of XCrySDen \cite{xcrysden}. \texttt{povray} stands for a scripting language used by the ray-tracing program POV-Ray. The core of the library consists of files \texttt{common.py}, \texttt{matrix.py}, and \texttt{matter.py}. Script \texttt{convert.py} is a command-line driver that performs the basic operations supported by the library. Script \texttt{link.py} is a molecular assembler that can be used to rotate and link two molecules together. Script \texttt{gsphere.py} generates a real-space grid and a sphere of ${\bf G}$-vectors given lattice parameters and a kinetic-energy cutoff. This helps to estimate the number of unoccupied states needed in GW calculations. Script \texttt{average.py} takes an average of the scalar field on the faces or in the volume of the unit cell. This is used to determine the vacuum level in DFT calculations. Script \texttt{volume.py} converts the a3Dr file produced by \texttt{PARATEC} or \texttt{BerkeleyGW} to Gaussian Cube or XCrySDen XSF format. Each script requires a different set of command-line arguments. Running individual scripts without arguments displays a list of all possible command line arguments and a short description of each argument.

\subsection{Band-structure interpolation}

To plot the quasiparticle band-structure of a system, we provide two methods. The first, \texttt{sig2wan} is a utility that uses Wannier interpolation of the band-structure \cite{marzari97, ivo01} and is based on the \texttt{Wannier90} \cite{wannier90} package. This utility does not construct the Wannier functions; the user has to do that using the package used to construct the mean-field eigenfunctions, \texttt{PARATEC} or \texttt{Quantum ESPRESSO}. Once the Wannier functions have been constructed, this utility replaces the mean-field eigenvalues in the \texttt{wannier.eig} file (generated by \texttt{Wannier90}) with the quasiparticle eigenvalues. The user then can rerun the \texttt{Wannier90} executable to generate band-structures along arbitrary directions. Because one replaces the mean-field eigenvalues with quasiparticle eigenvalues, this approach does not work well if one replaces only some of the mean-field eigenvalues with quasiparticle ones for entangled bands. The second method for plotting quasiparticle band-structure, the \texttt{inteqp} utility, uses Eq. \ref{kpeig} to construct the band-structure along arbitrary directions. In this method, we interpolate directly the quasiparticle corrections, which are significantly smoother functions of ${\bf k}$ and $E$ than the quasiparticle eigenvalues themselves. Therefore, this method requires both the mean-field eigenvalues and eigenfunctions along the desired band-structure direction.

\subsection{Other}

We provide a general-purpose utility called \texttt{mf\_convert.x} which can convert between binary and ASCII formats of wavefunction, density, and exchange-correlation potential files. The real/complex flavor is determined by reading the file header, and if the utility is called through \texttt{mf\_convert\_wrapper.sh}, the binary/ASCII format is detected via the \texttt{grep} command and need not be specified. This converter is useful for moving such files between different platforms, since the binary files are more compact and the form read by the code, but are not necessarily portable between different platforms, whereas the ASCII files are.

The image-charge model (ICM) is implemented in a utility called \texttt{icm.x}, based on the Surface code. For a molecule weakly coupled to a metallic surface, the self-energy correction to a state can be well approximated by the sum of the self-energy correction of that state in the isolated molecule and an additional term due to screening from the metal \cite{neaton}. This screening term is modeled as the electrostatic energy of the charge density of the wavefunction and its induced image-charge distribution in the metal. Let the operator $R$ be a reflection across an image plane. Then
%\begin{widetext}
\begin{align}
\Delta \Sigma = \pm \frac{1}{2} \int \int \psi \left( {\bf r} \right) \psi^{*} \left( {\bf r'} \right)
\frac{1}{\left| {\bf r} - R {\bf r'} \right|}
\psi \left( {\bf r'} \right) \psi^{*} \left( {\bf r} \right) d{\bf r} d{\bf r'}
\end{align}
%\end{widetext}
where the plus sign applies for occupied orbitals and the minus sign for unoccupied orbitals.
This approximation is useful for modeling scanning-tunneling spectroscopy of molecules absorbed on metal surfaces \cite{tao09} and for quantum-transport calculations of molecular junctions \cite{quek}.

\section{Acknowledgments}

We acknowledge the following people for their contributions to earlier version of the package: Xavier Blase, Andrew Canning, Eric K. Chang, Mark S. Hybertsen, Sohrab Ismail-Beigi, Je-Luen Li, Jeff Neaton, Cheol-Hwan Park, Filipe J. Ribeiro, Gian-Marco Rignanese, Catalin D. Spataru, Murilo L. Tiago, Li Yang and Peihong Zhang. We'd also like to thank the following beta users for their feedback, bug-reports, patches and general assistance while developing the code: Sangkook Choi, Peter Doak, Felipe Jornada, Brad D. Malone, Sahar Sharifzadeh, Isaac Tamblyn and Derek Vigil and the various other members of the Louie and Cohen groups at the University of California, Berkeley.

J.D. and M.J. acknowledge support from the Director, Office of Science, Office of Basic Energy Sciences, Materials Sciences and Engineering Division, U.S. Department of Energy under Contract No. DE-AC02-05CH11231. G.S. acknowledges support under National Science Foundation Grant No. DMR10-1006184. D.A.S. acknowledges support from the NSF Graduate Fellowship Program. Computational resources have been provided by NSF through TeraGrid resources at NICS and by DOE at Lawrence Berkeley National Laboratory's NERSC facility.

%%%%%%%%%%%%%%%%%%%%%%%%%%%%%%%%%%%%%%%%%%%%%%%%%%%%%%%%%%%%%%%
% reference
%%%%%%%%%%%%%%%%%%%%%%%%%%%%%%%%%%%%%%%%%%%%%%%%%%%%%%%%%%%%%%%%

\section{Appendix}

\subsection{Specification of file formats}

Wavefunction files are needed by all parts of the code, with various filenames. The \texttt{epsilon} executable uses an unshifted grid (\texttt{WFN}) and a shifted grid (\texttt{WFNq}). The \texttt{sigma} executable uses \texttt{WFN\_inner} to construct the self-energy operator and evaluates matrix elements with \texttt{WFN\_outer}. The \texttt{kernel} executable constructs kernel matrix elements with a coarse unshifted (\texttt{WFN\_co}) and shifted grid (\texttt{WFNq\_co}). The \texttt{absorption} and \texttt{plotxct} executables use a fine unshifted grid (\texttt{WFN\_fi}) and, with the velocity operator, a fine shifted grid (\texttt{WFNq\_fi}). Additionally, the \texttt{sigma} executable needs the charge-density \texttt{RHO} for GPP calculations, and needs the exchange-correlation potential \texttt{VXC} (unless its pre-computed matrix elements are supplied in a \texttt{vxc.dat} file). These files all share a common format, which begins with a header. Parts in italics are only for wavefunction files, not charge-density or exchange-correlation potential files. Each bullet represents a record in the file. The utility \texttt{wfn\_rho\_vxc\_info.x} can read the information from the header and report it in a comprehensible format to the user. A module of driver read/write routines for the formats specified here is available in the \texttt{library} directory, for mean-field codes to use in writing output for \texttt{BerkeleyGW}. This library is used by \texttt{Octopus} and \texttt{PARATEC}.

%\begin{widetext}
\begin{itemize}

\item {\tt [WFN/RHO/VXC]-[Real/Complex] date time

\item number of spins, number of G-vectors, number of symmetries, [0 for cubic symmetry/1 for hexagonal symmetry], number of atoms, charge-density cutoff (Ry), \textit{number of k-points, number of bands, maximum number of G-vectors for any k-point, wavefunction cutoff (Ry)}

\item FFT grid(1:3), \textit{k-grid(1:3), k-shift(1:3)}

\item real-space cell volume (a.u.), lattice constant (a.u.), lattice vectors(1:3, 1:3) in units of lattice constant, real-space metric tensor(1:3, 1:3) (a.u.)

\item reciprocal-space cell volume (a.u.), reciprocal lattice constant (a.u.), reciprocal lattice vectors(1:3, 1:3) in units of reciprocal lattice constant, reciprocal-space metric tensor(1:3, 1:3) (a.u.)

\item symmetry rotation matrices(1:3, 1:3, 1:number of symmetries) in reciprocal-lattice basis

\item symmetry fractional translations(1:3, 1:number of symmetries) in units of lattice vectors times $2 \pi$ (see also Sec. \ref{sec:symmetry_mf})

\item atomic positions(1:3, 1:number of atoms) in units of lattice constant, atomic numbers(1:number of atoms)

\item \textit{number of G-vectors for each k-point(1:number of k-points)}

\item \textit{k-point weights(1:number of k-points) from 0 to 1}

\item \textit{k-point coordinates(1:3, 1: number of k-points) in crystal coordinates}

\item \textit{index of lowest band to use on each k-point(1:number of k-points)}

\item \textit{index of highest occupied band on each k-point(1:number of k-points)}

\item \textit{energy eigenvalues(1:number of bands, 1:number of k-points, 1:number of spins) (Ry)}

\item \textit{occupations(1:number of bands, 1:number of k-points, 1:number of spins) from 0 to 1}

} \end{itemize} %\end{widetext}

In the body of a file, ${\bf G}$-vectors are listed as (1:3, 1:ng), expressed as integers in reciprocal lattice units, and data is listed as (1:ng, 1:number of spins). ${\bf G}$-vector components should be chosen in the interval $[-n/2, n/2)$ where $n$ is the FFT grid. A full sphere must be used, not a half sphere as in the Hermitian FFT representation for a real function. Each set is preceded by an integer specifying how many records the ${\bf G}$-vectors or data is broken up into, for ease of writing files from a code parallelized over ${\bf G}$-vectors. Wavefunction files follow the header with a listing of all the ${\bf G}$-vectors; for each ${\bf k}$-point, there is first a list of ${\bf G}$-vectors, and then the wavefunction coefficients for each band. \texttt{RHO} and \texttt{VXC} files have instead just one listing of ${\bf G}$-vectors and coefficients after the header. The wavefunction coefficients must be normalized so that the sum of their squares is 1. 
The \texttt{RHO} coefficients are normalized such that 
their ${\bf G} = 0$ component is the number of electrons 
in the unit cell. The \texttt{VXC} coefficients are in Ry.

The recommended scheme is to use pre-computed exchange matrix elements in an ASCII file \texttt{vxc.dat}, 
because \texttt{VXC} is only applicable to a local exchange-correlation functional. Hartree-Fock, 
hybrid-density functionals and self-consistent static GW (COHSEX) calculations do not fall in this category. 
For Hartree-Fock and some hybrid functionals (PBE0, B3LYP), one can still use \texttt{VXC} if one sets 
\texttt{bare\_exchange\_fraction} in \texttt{sigma.inp} to compensate for a fraction of the bare exchange 
which is not included in \texttt{VXC}. 
Matrix elements are in eV and are always written with real and imaginary parts (even in the real version of the code). 
The \texttt{vxc.dat} file may contain any number of ${\bf k}$-points in any order. It contains a certain number 
of diagonal elements (\texttt{ndiag}) and a certain number of offdiagonal elements (\texttt{noffdiag}). 
The \texttt{x.dat} file shares the same format and can supply saved matrix elements of bare exchange 
instead of calculating them. Each ${\bf k}$-point block begins with the line:

\texttt{kx, ky, kz [crystal coordinates], ndiag*nspin, noffdiag*nspin}

\noindent There are then \texttt{ndiag*nspin} lines of the form

\texttt{ispin, idiag, Re $\langle$idiag$|$V$|$idiag$\rangle$, Im $\langle$idiag$|$V$|$idiag$\rangle$}

\noindent There are then \texttt{noffdiag*nspin} lines of the form

\texttt{ispin, ioff1, ioff2, Re $\langle$ioff1$|$V$|$ioff2$\rangle$, Im $\langle$ioff1$|$V$|$ioff2$\rangle$}


\begin{thebibliography}{10}

\bibliographystyle{elsarticle-num}

\expandafter\ifx\csname url\endcsname\relax
  \def\url#1{\texttt{#1}}\fi
\expandafter\ifx\csname urlprefix\endcsname\relax\def\urlprefix{URL }\fi
\expandafter\ifx\csname href\endcsname\relax
  \def\href#1#2{#2} \def\path#1{#1}\fi

\bibitem{hybertsen86}
M.~S. Hybertsen, S.~G. Louie, Electron correlation in semiconductors and
  insulators: Band gaps and quasiparticle energies, Phys. Rev. B 34 (1986)
  5390.

\bibitem{louiechapter06}
S.~G. Louie, Predicting materials and properties: Theory of the ground and
  excited state, in: S.~G. Louie, M.~L. Cohen (Eds.), Conceptual Foundations of
  Materials: A Standard Model for Ground- and Excited-State Properties,
  Contemporary Concepts of Condensed Matter Science, Elsevier, Amsterdam, 2006,
  p.~9.

\bibitem{spataru04}
C.~D. Spataru, S.~Ismail-Beigi, L.~X. Benedict, S.~G. Louie, Excitonic effects
  and optical spectra of single-walled carbon nanotubes, Phys. Rev. Lett. 92
  (2004) 077402.

\bibitem{spataru04long}
C.~D. Spataru, S.~Ismail-Beigi, L.~X. Benedict, S.~G. Louie, Quasiparticle
  energies, excitonic effects and optical absorption spectra of small-diameter
  single-walled carbon nanotubes, Appl. Phys. A 78 (2004) 1129.

\bibitem{deslippe07}
J.~Deslippe, C.~D. Spataru, D.~Prendergast, S.~G. Louie, Bound excitons in
  metallic single-walled carbon nanotubes, Nano Lett. 7 (2007) 1626.

\bibitem{hedin65}
L.~Hedin, New method for calculating the one-particle {Green's} function with
  application to the electron-gas problem, Phys. Rev. 139~(3A) (1965)
  A796--A823.

\bibitem{strinati88}
G.~Strinati, Application of the {Green's} functions method to the study of the
  optical properties of semiconductors, Riv. Nuovo Cimento 11 (1988) 1.

\bibitem{rohlfing00}
M.~Rohlfing, S.~G. Louie, Electron-hole excitations and optical spectra from
  first principles, Phys. Rev. B 62 (2000) 4927.

\bibitem{reining98}
S.~Albrecht, L.~Reining, R.~Del~Sole, G.~Onida, Ab initio calculation of
  excitonic effects in the optical spectra of semiconductors, Phys. Rev. Lett.
  80~(20) (1998) 4510--4513.

\bibitem{benedict98}
L.~X. Benedict, E.~L. Shirley, R.~B. Bohn, Optical absorption of insulators and
  the electron-hole interaction: An ab initio calculation, Phys. Rev. Lett.
  80~(20) (1998) 4514--4517.

\bibitem{paratec}
\url{http://www.nersc.gov/projects/paratec/}.

\bibitem{espresso}
P.~Giannozzi, S.~Baroni, N.~Bonini, M.~Calandra, R.~Car, C.~Cavazzoni,
  D.~Ceresoli, G.~L. Chiarotti, M.~Cococcioni, I.~Dabo, A.~D. Corso,
  S.~de~Gironcoli, S.~Fabris, G.~Fratesi, R.~Gebauer, U.~Gerstmann,
  C.~Gougoussis, A.~Kokalj, M.~Lazzeri, L.~Martin-Samos, N.~Marzari, F.~Mauri,
  R.~Mazzarello, S.~Paolini, A.~Pasquarello, L.~Paulatto, C.~Sbraccia,
  S.~Scandolo, G.~Sclauzero, A.~P. Seitsonen, A.~Smogunov, P.~Umari, R.~M.
  Wentzcovitch, {QUANTUM ESPRESSO}: a modular and open-source software project
  for quantum simulations of materials, J. Phys.: Condens. Matt. 21~(39) (2009)
  395502.

\bibitem{soler02siesta}
J.~M. Soler, E.~Artacho, J.~D. Gale, A.~Garc\'ia, J.~Junquera, P.~Ordej\'on,
  D.~S\'anchez-Portal, The {\sc siesta} method for ab initio order-${N}$
  materials simulation, J. Phys.: Condens. Matt. 14 (2002) 2745.

\bibitem{parsec}
J.~R. Chelikowsky, N.~Troullier, Y.~Saad, Finite-difference-pseudopotential
  method: Electronic structure calculations without a basis, Phys. Rev. Lett.
  72~(8) (1994) 1240--1243.

\bibitem{parsec2}
M.~M.~G. Alemany, M.~Jain, L.~Kronik, J.~R. Chelikowsky, Real-space
  pseudopotential method for computing the electronic properties of periodic
  systems, Phys. Rev. B 69 (2004) 075101.

\bibitem{octopus_pssb}
A.~Castro, H.~Appel, M.~Oliveira, C.~A. Rozzi, X.~Andrade, F.~Lorenzen,
  M.~A.~L. Marques, E.~K.~U. Gross, A.~Rubio, \textit{octopus}: a tool for the
  application of time-dependent density functional theory, Phys. Status Solidi
  B 243~(11) (2006) 2465--2488.

\bibitem{octopus_CPC}
M.~A.~L. Marques, A.~Castro, G.~F. Bertsch, A.~Rubio, octopus: a
  first-principles tool for excited electron-ion dynamics, Comput. Phys.
  Commun. 151~(1) (2003) 60 -- 78.

\bibitem{martin04}
R.~M. Martin, Electronic Structure: Basic Theory and Practical Methods,
  Cambridge University Press, 2004.

\bibitem{kohn65}
W.~Kohn, L.~J. Sham, Self-consistent equations including exchange and
  correlation effects, Phys. Rev. 140 (1965) A1133.

\bibitem{perdew96}
J.~P. Perdew, K.~Burke, M.~Ernzerhof, Generalized gradient approximation made
  simple, Phys. Rev. Lett. 77 (1996) 3865.

\bibitem{hedin70}
L.~Hedin, S.~Lundqvist, Effects of electron-electron and electron-phonon
  interactions on the one-electron states of solids, in: F.~Seitz, D.~Turnbull,
  H.~Ehrenreich (Eds.), Advances in Research and Applications, Vol.~23 of Solid
  State Physics, Academic Press, 1970, pp. 1 -- 181.

\bibitem{cohen75}
M.~L. Cohen, M.~Schl{\"u}ter, J.~R. Chelikowsky, S.~G. Louie, Self-consistent
  pseudopotential method for localized configurations: Molecules, Phys. Rev. B
  12 (1975) 5575.

\bibitem{fetter}
A.~L. Fetter, J.~D. Walecka, Quantum Theory of Many-Body Systems, McGraw Hill,
  San Francisco, 1971.

\bibitem{jellison}
G.~E. Jellison, Jr., M.~F. Chisholm, S.~M. Gorbatkin, Optical functions of
  chemical vapor deposited thin-film silicon determined by spectroscopic
  ellipsometry, Appl. Phys. Lett. 62~(25) (1993) 3348--3350.

\bibitem{Holm98}
B.~Holm, U.~von Barth, Fully self-consistent ${GW}$ self-energy of the electron
  gas, Phys. Rev. B 57~(4) (1998) 2108--2117.

\bibitem{Aryasetiawan98}
F.~Aryasetiawan, O.~Gunnarsson, The ${GW}$ method, Rep. Prog. Phys. 61~(3)
  (1998) 237.

\bibitem{mjain10COHSEX}
M.~Jain, J.~Deslippe, G.~Samsonidze, M.~L. Cohen, S.~G. Louie, ${G}_0{W}_0$
  diagonalization using the static {COHSEX} approximation, Phys. Rev. B (to be
  published).

\bibitem{fleszarthesis}
A.~Fleszar, Dielectric response in semiconductors: theory and applications,
  Ph.D. thesis, University of Trieste (1985).

\bibitem{fftw}
M.~Frigo, S.~G. Johnson, The design and implementation of {FFTW3}, Proceedings
  of the IEEE 93~(2) (2005) 216--231, special issue on ``Program Generation,
  Optimization, and Platform Adaptation''.

\bibitem{blas}
E.~Anderson, Z.~Bai, C.~Bischof, S.~Blackford, J.~Demmel, J.~Dongarra,
  J.~Du~Croz, A.~Greenbaum, S.~Hammarling, A.~McKenney, D.~Sorensen, {LAPACK}
  Users' Guide, 3rd Edition, Society for Industrial and Applied Mathematics,
  Philadelphia, PA, 1999.

\bibitem{baldereschi78}
A.~Baldereschi, E.~Tosatti, Mean-value point and dielectric properties of
  semiconductors and insulators, Phys. Rev. B 17~(12) (1978) 4710--4717.

\bibitem{hightemp}
L.~X. Benedict, C.~D. Spataru, S.~G. Louie, Quasiparticle properties of a
  simple metal at high electron temperatures, Phys. Rev. B 66~(8) (2002)
  085116.

\bibitem{sahar}
S.~Sharifzadeh, A.~Biller, L.~Kronik, J.~B. Neaton, Quasiparticle and optical
  spectroscopy of the organic semiconductors pentacene and {PTCDA} from first
  principles, Phys. Rev. B 85 (2012) 125307.

\bibitem{shih10}
B.-C. Shih, Y.~Xue, P.~Zhang, M.~L. Cohen, S.~G. Louie, Quasiparticle band gap
  of {ZnO}: High accuracy from the conventional ${G}_0{W}_0$ approach, Phys.
  Rev. Lett. 105 (2010) 146401.

\bibitem{rinke05}
P.~Rinke, A.~Qteish, J.~Neugebauer, C.~Freysoldt, M.~Scheffler, Combining
  ${GW}$ calculations with exact-exchange density-functional theory: an
  analysis of valence-band photoemission for compound semiconductors, New J.
  Phys. 7 (2005) 126.

\bibitem{bruneval06}
F.~Bruneval, N.~Vast, L.~Reining, Effect of self-consistency on quasiparticles
  in solids, Phys. Rev. B 74 (2006) 045102.

\bibitem{vanschilfgaarde06}
M.~van Schilfgaarde, T.~Kotani, S.~Faleev, Quasiparticle self-consistent ${GW}$
  theory, Phys. Rev. Lett. 96 (2006) 226402.

\bibitem{spataruthesis}
C.-D. Spataru, Electron excitations in solids and novel materials, Ph.D.
  thesis, University of California, Berkeley (2004).

\bibitem{zhang89}
S.~B. Zhang, D.~Tom{\'a}nek, M.~L. Cohen, S.~G. Louie, M.~S. Hybertsen,
  Evaluation of quasiparticle energies for semiconductors without inversion
  symmetry, Phys. Rev. B 40 (1989) 3162.

\bibitem{hanke78}
W.~Hanke, Dielectric theory of elementary excitations in crystals, Adv. Phys.
  27~(2) (1978) 287--341.

\bibitem{haydock80}
R.~Haydock, The recursive solution of the {Schr\"{o}dinger }equation, Comput.
  Phys. Commun. 20~(1) (1980) 11 -- 16.

\bibitem{benedict99}
L.~X. Benedict, E.~L. Shirley, Ab initio calculation of $\epsilon_2(\omega)$
  including the electron-hole interaction: Application to {GaN} and {CaF}$_2$,
  Phys. Rev. B 59~(8) (1999) 5441--5451.

\bibitem{tiagothesis}
M.~L. Tiago, Electronic and optical properties of organic crystals, polymers,
  and semiconductors, Ph.D. thesis, University of California, Berkeley (2003).

\bibitem{sohrab01}
S.~Ismail-Beigi, E.~K. Chang, S.~G. Louie, Coupling of nonlocal potentials to
  electromagnetic fields, Phys. Rev. Lett. 87 (2001) 087402.

\bibitem{scalapack}
L.~S. Blackford, J.~Choi, A.~Cleary, E.~D'Azevedo, J.~Demmel, I.~Dhillon,
  J.~Dongarra, S.~Hammarling, G.~Henry, A.~Petitet, K.~Stanley, D.~Walker,
  R.~C. Whaley, {ScaLAPACK} Users' Guide, Society for Industrial and Applied
  Mathematics, Philadelphia, PA, 1997.

\bibitem{yang09}
L.~Yang, J.~Deslippe, C.-H. Park, M.~L. Cohen, S.~G. Louie, Excitonic effects
  on the optical response of graphene and bilayer graphene, Phys. Rev. Lett.
  103~(18) (2009) 186802.

\bibitem{beigi06}
S.~Ismail-Beigi, Truncation of periodic image interactions for confined
  systems, Phys. Rev. B 73 (2006) 233103.

\bibitem{spglib}
A.~Togo, \url{http://spglib.sourceforge.net/}.

\bibitem{hybertsen87}
M.~S. Hybertsen, S.~G. Louie, Ab initio static dielectric matrices from the
  density-functional approach. {I}. {Formulation} and application to
  semiconductors and insulators, Phys. Rev. B 35~(11) (1987) 5585--5601.

\bibitem{gpzhang}
G.~P. Zhang, D.~A. Strubbe, S.~G. Louie, T.~F. George, First-principles
  prediction of optical second-order harmonic generation in the endohedral
  {N}@{C}$_{60}$ compound, Phys. Rev. A 84~(2) (2011) 023837.

\bibitem{sciprog}
Z.~Mirali, ...{ERROR}...why scientific programming does not compute, Nature 467
  (2010) 775 -- 777.

\bibitem{SVN}
\url{http://subversion.tigris.org/}.

\bibitem{Trac}
\url{http://trac.edgewall.org/}.

\bibitem{BuildBot}
\url{http://buildbot.net/}.

\bibitem{povray}
\url{http://www.povray.org/}.

\bibitem{xcrysden}
A.~Kokalj, \href{http://www.xcrysden.org/}{Computer graphics and graphical user
  interfaces as tools in simulations of matter at the atomic scale}, Comp.
  Mater. Sci. 28 (2003) 155.
\newline\urlprefix\url{http://www.xcrysden.org/}

\bibitem{marzari97}
N.~Marzari, D.~Vanderbilt, Maximally localized generalized {Wannier} functions
  for composite energy bands, Phys. Rev. B 56~(20) (1997) 12847--12865.

\bibitem{ivo01}
I.~Souza, N.~Marzari, D.~Vanderbilt, Maximally localized {Wannier} functions
  for entangled energy bands, Phys. Rev. B 65~(3) (2001) 035109.

\bibitem{wannier90}
A.~A. Mostofi, J.~R. Yates, Y.-S. Lee, I.~Souza, D.~Vanderbilt, N.~Marzari,
  wannier90: A tool for obtaining maximally-localised {Wannier} functions,
  Comput. Phys. Commun. 178~(9) (2008) 685 -- 699.

\bibitem{neaton}
J.~B. Neaton, M.~S. Hybertsen, S.~G. Louie, Renormalization of molecular
  electronic levels at metal-molecule interfaces, Phys. Rev. Lett. 97~(21)
  (2006) 216405.

\bibitem{tao09}
C.~Tao, J.~Sun, X.~Zhang, R.~Yamachika, D.~Wegner, Y.~Bahri, G.~Samsonidze,
  M.~L. Cohen, S.~G. Louie, T.~D. Tilley, R.~A. Segalman, M.~F. Crommie,
  Spatial resolution of a type {II} heterojunction in a single bipolar
  molecule, Nano Lett. 9 (2009) 3963.

\bibitem{quek}
S.~Y. Quek, D.~A. Strubbe, H.~J. Choi, S.~G. Louie, J.~B. Neaton,
  First-principles approach to charge transport in single-molecule junctions
  with self-energy corrections: a {DFT}+{$\Sigma$} method, in preparation.

\end{thebibliography}
\end{document}